\documentclass[aps,pre,twocolumn]{revtex4-1}

\usepackage{amsmath}
\usepackage{amssymb}
\usepackage{color}
\usepackage{hyperref}
\usepackage{bm}
\usepackage[normalem]{ulem}

\usepackage{graphicx}

\DeclareMathAlphabet{\mathitbf}{OML}{cmm}{b}{it}

\renewcommand{\th}{^{\mbox{\tiny th}}}
\renewcommand{\=}{\!=\!}

\newcommand{\dbar}{{\,\mathchar'26\mkern-12mu d}}

\newcommand{\sFrac}[2]{{\textstyle\frac{#1}{#2}}}
\newcommand{\tripleCdot}{\stackrel{\mbox{\bf\scriptsize .}}{:}}

\begin{document}
\title{A characteristic energy scale in glasses}
\author{Edan Lerner${}^{1}$ and Eran Bouchbinder${}^{2}$}
\affiliation{${}^1$Institute for Theoretical Physics, University of Amsterdam, Science Park 904, 1098 XH Amsterdam, The Netherlands \\ ${}^2$Chemical and Biological Physics Department, Weizmann Institute of Science, Rehovot 7610001, Israel}

\begin{abstract}
Intrinsically generated structural disorder endows glassy materials with a broad distribution of various microscopic quantities --- such as relaxation times and activation energies --- without an obvious characteristic scale. At the same time, macroscopic glassy response --- such as Newtonian (linear) viscosity and nonlinear plastic deformation --- is widely interpreted in terms of a characteristic energy scale, e.g.~an effective temperature-dependent activation energy in Arrhenius relations. Nevertheless, despite its fundamental importance, such a characteristic energy scale has not been robustly identified. Inspired by the accumulated evidence regarding the crucial role played by disorder- and frustration-induced soft quasilocalized excitations in determining the properties and dynamics of glasses, we propose that the bulk average of the glass response to a localized force dipole defines such a characteristic energy scale. We show that this characteristic glassy energy scale features remarkable properties: $(i)$ It increases dramatically in underlying inherent structures of equilibrium supercooled states approaching the glass transition temperature $T_g$, significantly surpassing the corresponding increase in the macroscopic shear modulus, dismissing the common view that structural variations in supercooled liquids upon vitrification are minute $(ii)$ Its variation with annealing and system size is very similar in magnitude and form to that of the energy of the softest non-phononic vibrational mode, thus establishing a nontrivial relation between a rare glassy fluctuation and a bulk average response $(iii)$ It exhibits striking dependence on spatial dimensionality and system size, due to the long-ranged fields associated with quasilocalization, which are speculated to be related to peculiarities of the glass transition in two dimensions. In addition, we identify a truly-static growing lengthscale associated with the characteristic glassy energy scale, and discuss possible connections between the increase of this energy scale and the slowing down of dynamics near the glass transition temperature. Open questions and future directions are discussed.
\end{abstract}

\maketitle

\section{introduction}
\label{introduction}

The dynamical processes accompanying the formation and deformation of glassy materials continue to pose perplexing riddles even after several decades of research~\cite{jeppe_review2006,heuer_review,Cavagna200951, Berthier_Biroli_RevModPhys_2011, paddy_huge_review_2015, falk_review_2016,  charbonneau_review_2017, bonn_yield_stress_review_2017}. One influential viewpoint on thermally-activated or externally-driven glassy dynamics is known as the potential energy landscape (PEL) picture, put forward first by Goldstein~\cite{Goldstein1969} in the context of supercooled liquids' dynamics. According to this viewpoint, the coordinates $\vec{x}$ of a glassy configuration are envisioned as a state point on the multi-dimensional surface (landscape) as defined by the potential energy $U(\vec{x})$, which is expected to be highly rugged due to the presence of structural disorder.

The onset of slow dynamics in supercooled liquids within the PEL picture is associated with the occurrence of a crossover in the typical environments of the PEL sampled by the system as temperature is lowered~\cite{SASTRY1999301,landscape_dominated_jeppe_2000,Cavagna_prl_2000,Cavagna_prl_2002}. In particular, below some crossover temperature, the system is assumed to reside near local minima of the highly-rugged multi-dimensional PEL, such that structural relaxation --- the motion of the state point into the basins of other, distant local minima --- requires the occurrence of activated processes over potential energy barriers that surround local minima. From this perspective, the increase of the relaxation time with deeper supercooling might be interpreted in terms of changes in the magnitude and number of accessible potential energy barriers~\cite{jeppe_review2006,Cavagna200951}. The distribution of relaxation times and energy barriers is expected to be broad, reflecting the disordered nature of glassy systems.

The PEL picture has also guided investigations of the yielding transition in driven systems, that inevitably occurs when a glassy solid is subjected to sufficiently large external forces. Within the PEL viewpoint, the micromechanical processes that give rise to irreversible plastic flow in glassy solids correspond to the deformation-induced development of soft or even unstable directions on the PEL, that allow the system to escape the vicinity of a local minimum and flow towards the basins of others~\cite{Malandro_Lacks,lemaitre2004,micromechanics2016}. Information about the susceptibility of a glassy solid to deform plastically is therefore presumably encoded in its PEL properties. There is currently no general agreement, however, about how those properties are best defined and probed~\cite{tanguy2010,manning2011,rottler_normal_modes,manning_defects,plastic_modes_prerc,thermal_energies,machine_learning_1,machine_learning_3}.

Despite its conceptual simplicity, this appealing picture of thermally-activated and externally-driven glassy dynamics is difficult to test directly due to both the multi-dimensional and highly-rugged nature of the PEL~\cite{wales2003energy, Schuh_prl_2009, Rodney2011}. Complementary real-space numerical studies of the dynamics of supercooled liquids~\cite{Schober_correlate_modes_dynamics,landscape_dominated_jeppe_2000,Cavagna_prl_2002,kob_2006_compact_cluster_motion,Chandler_prx,lemaitre_scale_free_prl} and of driven glassy solids~\cite{falk_langer_stz,Malandro_Lacks,lemaitre2004} clearly indicate that relaxation/flow proceeds via spatially localized, predominantly shear-like rearrangements of a few tens of particles. These observations suggest that although the highly-rugged PEL of generic glass-formers features a broad distribution of activation barriers, only a particular --- and apparently very small --- sub-class of excursions from the vicinity of local minima of the PEL  --- that correspond to localized rearrangements in real-space --- are relevant for structural relaxation or plastic flow.

How can directions in the highly-rugged multi-dimensional PEL that are particularly relevant for relaxation and flow be characterized? In turn, this question admits a real-space representation: what characterization of \emph{local microstructures} in glassy materials is indicative of susceptibility to relaxation and flow? In this work we put forward and carefully test the proposition that the spatial response of glassy materials to local force dipoles reveals the directions in configuration-space that are relevant to glassy dynamics, and exposes the relevant characteristic energy scale for thermally-activated or externally-driven relaxation processes.

Our work is focused on studying three major aspects of the characteristic energy scale defined by the average response to local force dipoles, referred to in what follows as the \emph{characteristic glassy energy scale} (CGE). First, we examine how the CGE varies in inherent states (local minima of the potential energy) that underlie equilibrium supercooled liquid configurations at various temperatures, devoting particular attention to its variation as the glass transition temperature $T_g$ is approached. Remarkably, we find that the CGE changes by almost $100\%$ between ensembles of glassy samples generated by instantaneous quenches from different equilibrium parent temperatures $T_0$. This large variation is meaningfully compared to e.g.~the stiffening of the shear modulus, which shows a relative variation of merely $\sim\!25\%$ across the same range of $T_0$.

We also investigate theoretically and numerically how the CGE depends on spatial dimensionality and on system size. While we find negligible dependence of the CGE on system size in three-dimensions (3D), the situation in two-dimensions (2D) is dramatically different: we find a factor of almost 3 in the relative variation of the CGE over simulationally-accessible system sizes in 2D, and predict that it vanishes logarithmically in the thermodynamic limit.

Finally, in both aforementioned parts of the current study, we also report concrete evidence that the CGE is an excellent representative of the typical energies of soft, quasilocalized excitations --- the `softest spots' in the material. The relation we establish is nontrivial: we show that the energy scale of the rarest structural fluctuations is closely related to a bulk average response, namely the CGE. This leads us to propose and test a relation between the variation in activation barriers toward structural relaxation in deeply supercooled liquid states, and the observed increase of the CGE in the underlying inherent states.

This paper is organized as follows; in Sect.~\ref{characteristic_glassy_energy} we directly motivate, define and discuss the central observable of our work, the CGE; in particular, we explain why it is proposed to characterize the directions in configuration-space that are most relevant to glassy dynamics and the corresponding spatially-localized regions in the glass, being inspired by the accumulated evidence for the importance of soft quasilocalized excitations for glassy dynamics. In Sect.~\ref{protocol_section} we study how the CGE depend on the protocol used to create the glasses in which they are measured, and report results suggesting an intrinsic connection between the CGE and very soft quasilocalized excitations. In Sect.~\ref{sec:connection} we firmly establish the suggested connection between the CGE and the energies of the softest quasilocalized excitations in glassy samples using their system size dependence. In Sect.~\ref{system_size_section} we focus on the spatial dimensionality and system size dependencies of the CGE. Finally, in Sect.~\ref{sec:relaxation} we explore the connection between the CGE and the slowing down of relaxational dynamics with decreasing temperature. Our findings and their implications are discussed in Sect.~\ref{discussion}, where several future research directions are proposed. We note that a complete description of the employed numerical glass-forming model, and of the protocols employed to generate ensembles of glassy samples, are provided in Appendix~\ref{appendix}. In what follows we omit the units of all reported observables; they should be understood as expressed in terms of the relevant microscopic units, as detailed in Appendix~\ref{appendix}.

\section{characteristic glassy energy scale}
\label{characteristic_glassy_energy}

Many studies of the mechanical properties of glassy solids and of the relaxational dynamics of supercooled liquids are focused on establishing causal structure-dynamics relations~\cite{harrowell_isoconfiguration, rob_jack_information_prl_2014, glen_prl_2014, paddy_huge_review_2015, Coslovich_2007_lfs,turci_prx_2017, royall_prl_2017_experiment_lfs, tanaka_prl_2007,Leocmach2012, tanaka_2006, tah2017glass, Montanari2006,smarajit_review, Schober_correlate_modes_dynamics, wyart_brito_2007,widmer2008irreversible,harrowell_2009, manning2011, rottler_normal_modes, manning_defects, thermal_energies, plastic_modes_prerc, micromechanics2016, machine_learning_1, falk_prl_2016, machine_learning_2, machine_learning_3}. This is a remarkably challenging task given the well-known striking absence of structural variations that are able to simply explain the vast variations in rates of dynamical processes. Two conventional examples of this infamous hallmark of glassy dynamics, in relation to both the temperature dependence of the primary equilibrium structural relaxation time and the effect of the quenching rate on the nonlinear mechanical response, are shown in Appendix~\ref{sec:structure_dynamics}. In both cases, dramatic dynamical effects --- e.g.~a relative increase of 150\% in the free-energy activation barriers for structural relaxation or huge differences in the yielding dynamics --- are accompanied by~very minor changes in pair correlation functions (see Appendix~\ref{sec:structure_dynamics}).


There are two main approaches to establishing causal structure-dynamics relations in equilibrium supercooled liquids or deformed glassy solids: one that focuses on the identification of so-called locally-favored structures~\cite{paddy_huge_review_2015, glen_prl_2014, Coslovich_2007_lfs,turci_prx_2017, royall_prl_2017_experiment_lfs, tanaka_prl_2007,Leocmach2012, tanaka_2006, tah2017glass, Montanari2006,smarajit_review} and one that focuses on the identification of some sort of `flow-defects', or `soft spots' --- localized regions in glassy samples that are particularly susceptible to plastic rearrangements under external deformations or to activated rearrangements under thermal fluctuations~\cite{Schober_correlate_modes_dynamics, wyart_brito_2007,widmer2008irreversible,harrowell_2009, manning2011, rottler_normal_modes, manning_defects, thermal_energies, plastic_modes_prerc, micromechanics2016, machine_learning_1, falk_prl_2016, machine_learning_2, machine_learning_3}. Among these two dual approaches, we follow the latter.

\begin{figure*}[!ht]
  \begin{minipage}[c]{0.60\textwidth}
    \includegraphics[width=\textwidth]{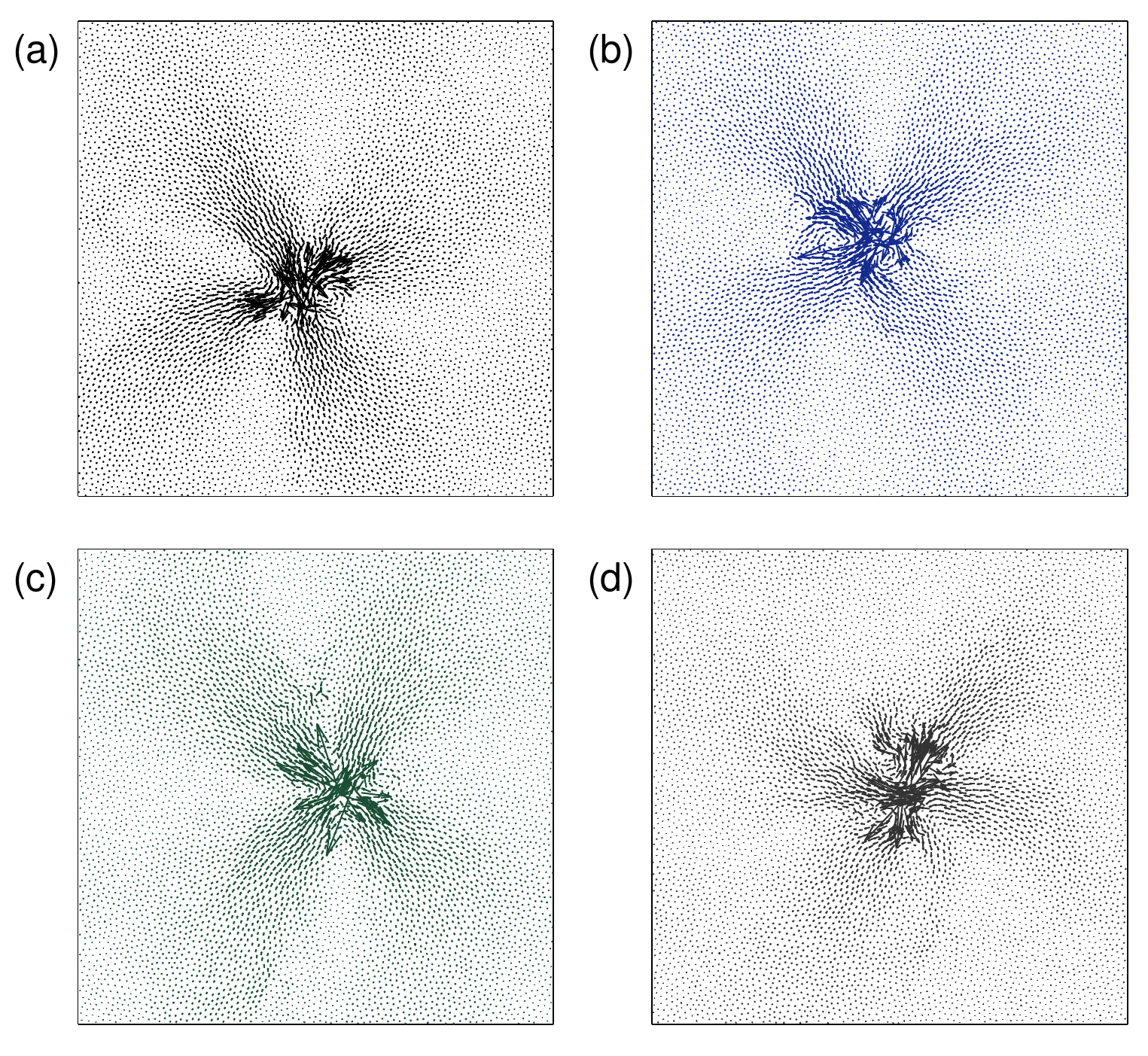}
  \end{minipage}\hfill
  \begin{minipage}[c]{0.35\textwidth}
    \caption{\footnotesize Various quasilocalized excitations (QLE) measured in different realizations of 2D computer glasses (see details of the model in the main text and in Appendix~\ref{appendix}): {\bf (a)} a low-frequency quasilocalized harmonic vibrational mode, {\bf (b)} a shear-transformation zone (plastic rearrangement) observed under simple shear deformation, {\bf (c)} a nonlinear glassy mode (see Sect.~\ref{system_size_section} of main text), and {\bf (d)} the displacement response to a local force dipole.}
 \label{introduction_fig}
  \end{minipage}
\end{figure*}

Soft, quasilocalized excitations (QLE) that are embedded in glasses' microstructure, as depicted for example in Fig.~\ref{introduction_fig}, are known to play a key role in glassy dynamics.  Several recent developments~\cite{modes_prl, inst_note, SciPost2016, thermal_energies,protocol_prerc,geert_prl,ikeda_pnas} have shed new light on the statistical and structural properties of such soft QLE. It is now broadly accepted that QLE emerge from the presence of frustration-induced internal stresses in the material \cite{inst_note}, and are observed to feature scale-free statistics~\cite{modes_prl, inst_note, SciPost2016, thermal_energies,protocol_prerc,geert_prl}.

Under some conditions, QLE assume the form of quasilocalized vibrational (harmonic) modes that dwell at the low frequency end of the vibrational spectrum of simple model glasses. This may occur in (i) systems made small enough to sufficiently suppress low-frequency phonons~\cite{modes_prl, SciPost2016, protocol_prerc, inst_note}, or (ii) systems with the precise size (as seen e.g.~in \cite{ikeda_pnas}) such that a coexistence frequency-window for quasilocalized vibrational modes and phonons opens, see \cite{phonon_widths} for an elaborate discussion on this matter. Under a broad range of circumstances, as discussed at length in \cite{protocol_prerc}, the distribution of quasilocalized vibrational modes of frequency $\omega$ in generic 3D computer models of structural glasses grows from zero as $\omega^4$, independently of microscopic details \cite{modes_prl,geert_prl}, as predicted decades ago within the Soft Potential Model framework \cite{soft_potential_model_1987,soft_potential_model_1991}, and more recently in \cite{Gurevich2003,Gurevich2007}. When conditions (i) or (ii) are not satisfied one needs to resort to nonlinear measures~\cite{SciPost2016} to reveal QLE, whose existence and statistics are independent of whether they can be realized as harmonic modes or not.

In view of the piling evidence for the central role played by QLE in glassy dynamics, we expect the vast variations observed in glassy dynamics to stem from changes in the characteristic energy scale associated with QLE. How can the characteristic scale of QLE be robustly identified? To the best of our knowledge, the answer to this important question is still unknown. Naively, a characteristic scale could be extracted from the density of harmonic vibrational modes whose low-frequency tail takes the form~\cite{modes_prl}
\begin{equation}
D(\omega) \propto \tilde{\omega}_g^{-5}\omega^4\,,
\end{equation}
where the relative variations of the protocol-dependent prefactor $\tilde{\omega}_g^{-5}$ could be interpreted as variations in QLE's characteristic scale. However, two immediate problems arise: first, many studies of the structural properties of glasses employ computer glasses created by instantaneous, overdamped quenches, for which $D(\omega)\!\sim\!\omega^\chi$ with $\chi\!<\!4$ is observed~\cite{protocol_prerc}, where $\chi$ depends delicately on the parent equilibrium temperature from which the quench is performed. This hinders a systematic comparison between scales deduced for glasses created via different protocols. Second, and more importantly, as there is no constraint whatsoever on the total \emph{number} of quasilocalized vibrational modes per unit volume, one should in fact generally write
\begin{equation}
D(\omega) \propto {\cal N}\omega_g^{-5}\omega^4\,,
\end{equation}
where ${\cal N}$ is a protocol-dependent normalization factor that fixes the total number of quasilocalized vibrational modes per unit volume, and $\omega_g^2$ is the sought-for protocol-dependent characteristic energy of QLE. An essentially identical argument was made in \cite{modes_prl} regarding the system-size dependence of the sample-to-sample statistics of the minimal vibrational frequency, which was concluded to depend on a protocol-dependent `site length', in addition to the characteristic scale $\omega_g$ and the system size. We conclude therefore that the prefactor of the low-frequency tails of the density of vibrational modes cannot reliably yield the desired characteristic energy scale.

\begin{figure*}[!ht]
\centering
\includegraphics[width = 0.95\textwidth]{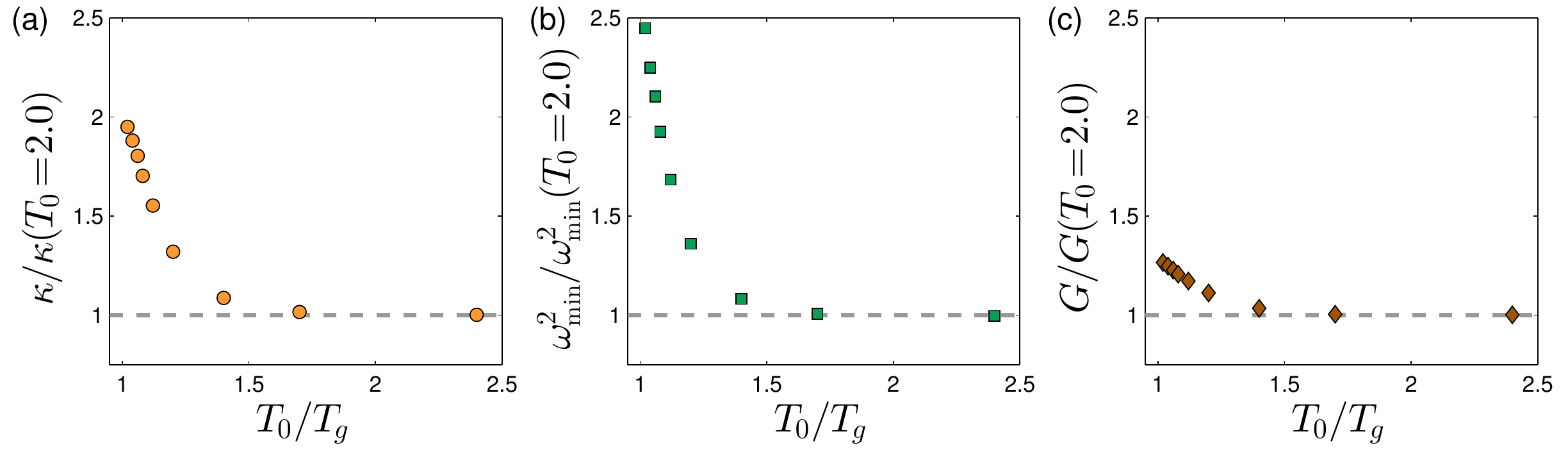}
\caption{\footnotesize Sample-to-sample averages of (a) the characteristic glassy energies $\kappa$, (b) the minimal vibrational frequency squared $\omega_{\mbox{\tiny min}}^2$, and (c) the athermal shear modulus $G$, rescaled by their high-$T_0$ limits, and plotted against the equilibrium parent temperature $T_0$ from which our ensembles of glassy samples were instantaneously quenched.}
\label{kappa_fig}
\end{figure*}

Here we propose instead that the bulk average of the glass response to a localized force dipole (see precise definition below) defines the characteristic energy scale $\omega_g^2$ of QLE. This proposition is motivated by the close resemblance between the spatial structure of the dipolar response of glasses and QLE, as demonstrated in Fig~\ref{introduction_fig}. All examples of QLE --- a low-frequency harmonic vibrational mode, a shear-transformation-zone (STZ) observed under external driving forces~\cite{lemaitre2004} and a nonlinear glassy mode~\cite{plastic_modes_prerc,micromechanics2016,SciPost2016} --- exhibit striking resemblance to the dipolar response: they all typically feature a disordered core decorated by Eshelby-like~\cite{Eshelby} (mostly-affine) fields that decay away from the core as $r^{1-\dbar}$ in $\dbar$ spatial dimensions (see additional discussion of these field in Sect.~\ref{system_size_section}).

To define the characteristic glassy energy scale studied in this work, consider a system of $N$ particles interacting via a radially-symmetric pairwise potential $\varphi(r)$, with $r$ the pairwise distance between particles. Labeling pairs of particles by $\alpha$, a force dipole on the $\alpha\th$ pair is defined as
\begin{equation}
\vec{d}_\alpha = \frac{\partial \varphi_\alpha}{\partial \vec{x}}\ ,
\end{equation}
where $\vec{x}$ denotes the vector of $N\!\times\!\dbar$ particles' coordinates in $\dbar$ dimensions.
$\vec{d}_\alpha$ is an $N\!\times\!\dbar$-dimensional vector which has non-vanishing components only at the $\alpha\th$ pair,
having the geometry of a dipole vector, i.e.~two forces of equal magnitude acting in opposite directions.

The linear response $\vec{z}_\alpha$ to an imposed force dipole on the $\alpha\th$ pair satisfies
\begin{equation}
\label{eq:dipolar_response}
{\cal M}\cdot\vec{z}_\alpha = \vec{d}_\alpha\,,
\end{equation}
where ${\cal M}\!\equiv\!\frac{\partial^2U}{\partial\vec{x}\partial\vec{x}}$ denotes the Hessian matrix of the potential energy $U\!=\!\sum_\alpha \varphi_\alpha(r_\alpha)$. An example of $\vec{z}_\alpha$ is displayed in Fig.~\ref{introduction_fig}d, where the striking spatial resemblance to QLE is demonstrated. This resemblance leads us to propose that $\vec{z}_\alpha$ picks up directions in the highly-rugged multi-dimensional PEL that are particularly relevant for glassy relaxation and flow. In particular, we are interested in the properties of the PEL in these special directions and in the associated characteristic energy scale. The goal of the remainder of the paper is to test this idea and its physical implications.

To this aim, we associate a stiffness $\kappa_\alpha$ with the response $\vec{z}_\alpha$ according to
\begin{equation}\label{foo00}
\kappa_\alpha \equiv \frac{\vec{z}_\alpha\cdot{\cal M}\cdot\vec{z}_\alpha}{\vec{z}_\alpha\cdot\vec{z}_\alpha} = \frac{\vec{d}_\alpha\cdot{\cal M}^{-1}\cdot\vec{d}_\alpha}{\vec{d}_\alpha\cdot{\cal M}^{-2}\cdot\vec{d}_\alpha}\,.
\end{equation}
As the energy of a unit displacement in the direction of $\vec{z}_\alpha$ is equal to $\case{1}{2}\kappa_\alpha$ in the harmonic approximation, we refer in what follows to $\kappa_\alpha$ as the {\em energy} associated with the glassy linear response to local force dipoles, keeping in mind that its actual units are energy per length squared. Finally, the characteristic glassy energy scale $\kappa$, referred to in what follows as the \emph{characteristic glassy energy scale} (CGE), is defined as the average of $\kappa_\alpha$ taken over all pairs of nearest neighbors. Taking the bulk average of $\kappa_\alpha$ is a non-trivial step; for example, it is not a priori clear that by so doing one does not wash out all relevant information about the softest glassy excitations in the system. This point will be extensively discussed below.

\section{Effect of annealing on the characteristic energy scale}
\label{protocol_section}

We turn now to investigate how the CGE defined by the response to local force dipoles depends on the protocol with which our computer glasses were created.  To this aim, we created two sets of ensembles of glassy samples: (i) ensembles of glassy samples that were quenched instantaneously from independent equilibrium configurations at a parent temperature $T_0$, and (ii) ensembles of glassy samples that were quenched from equilibrium at a finite quench rate $\dot{T}$. A complete description of our numerical protocols is available in Appendix~\ref{appendix}.

Our key findings for the instantaneously-quenched ensembles are displayed in Fig.~\ref{kappa_fig}; in panel (a) we plot the CGE $\kappa$ (obtained by averaging over both interactions and samples) vs.~the parent equilibrium temperature $T_0$ from which the glassy samples were instantaneously quenched. Remarkably, we find that the CGE increases significantly as the parent temperature approaches the glass transition temperature: the relative variation of $\kappa$ throughout the sampled parent-temperature range approaches a factor of $2$. This should be contrasted with the variation in the athermal shear modulus $G$, reported in Fig.~\ref{kappa_fig}c, which changes by merely 26\% throughout the entire parent-temperature range.

In the system size utilized, of $N\!=\!2000$ in 3D, the lowest vibrational frequency is always associated with a quasilocalized vibrational mode \cite{modes_prl}. In Fig.~\ref{kappa_fig}b we show the sample-to-sample averages of the minimal vibrational frequency squared $\omega_{\mbox{\tiny min}}^2$ vs.~the parent temperature $T_0$. Similarly to the CGE, this energy scale also changes significantly upon better annealing of the inherent states, with a relative variation that approaches a factor of $2.5$.
\begin{figure}[!ht]
\centering
\includegraphics[width = 0.50\textwidth]{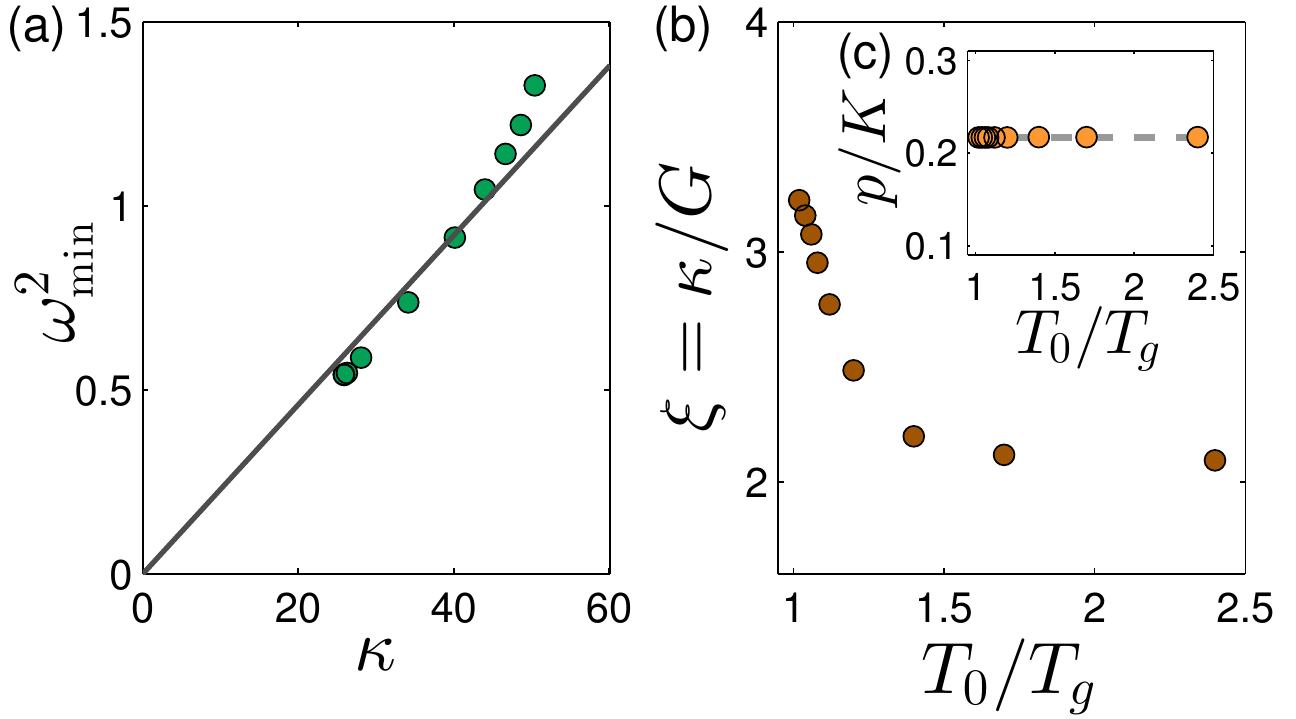}
\caption{\footnotesize (a) Sample-to-sample average CGEs $\kappa$ vs.~sample-to-sample average minimal vibrational frequency squared $\omega_{\mbox{\tiny min}}^2$. The continuous line is a guide to the eye. (b) static length $\xi\!\equiv\!\kappa/G$ extracted via the CGEs and shear moduli, vs.~parent temperature $T_0$. (c) Sample-to-sample average pressure to bulk modulus ratio, which shows no measureable variation upon annealing.}
\label{kappa_vs_omega_fig}
\end{figure}

In Fig.~\ref{kappa_vs_omega_fig}a we plot the minimal vibrational frequency squared $\omega_{\mbox{\tiny min}}^2$ against the CGE $\kappa$ and observe that while these two scales share the same units, two order of magnitude separate their measured values; this difference in magnitude is expected as the former is a rare fluctuation and the latter is a bulk average. Nevertheless, it is clear that the variations in these energy scales are highly correlated, though not strictly proportional, suggesting a nontrivial relation between a rare structural fluctuation, $\omega_{\mbox{\tiny min}}^2$, and a bulk average response function, $\kappa$. The lack of strict proportionality between $\omega_{\mbox{\tiny min}}^2$ and the CGE, as seen in Fig.~\ref{kappa_vs_omega_fig}a, underlines our assertion that a characteristic glassy energy scale cannot be directly extracted from the prefactor of the $\omega^4$ density of states. We propose that this lack of strict proportionality demonstrates that, in addition to variations in the typical energy of QLE upon annealing, their number per unit volume is also depleted.  Since $\omega_{\mbox{\tiny min}}^2$ in each glassy sample is the minimal amongst the energies of a population of QLE, one expects $\omega_{\mbox{\tiny min}}^2$ to increase if that population is depleted, resulting in a sharper increase of $\omega_{\mbox{\tiny min}}^2$ with better annealing compared to the increase in $\kappa$. 



Our analysis of the CGE and of the shear modulus in the instantaneously-quenched samples gives rise to the definition of a static lengthscale $\xi\!\equiv\!\kappa/G$, which is plotted against the parent temperature $T_0$ in Fig.~\ref{kappa_vs_omega_fig}b.  We note that this static lengthscale clearly does not describe the core size of QLE; in \cite{protocol_prerc} the latter was indirectly shown to decrease with deeper supercooling, whereas the static length $\xi(T_0)$ is clearly observed to increase with deeper supercooling. Possibly related lengths defined via the boson peak frequency and the speed of sound were put forward and measured experimentally by Sokolov and co-workers \cite{sokolov_boson_peak_scale}. In~\cite{karmakar_lengthscale,karmakar_lengthscale2,karmakar_lengthscale3} a closely-related effort to extract growing static lengths in annealed inherent states was put forward, based on the different variation with annealing of the Debye and QLE frequencies.

Within the unjamming scenario of packings of purely repulsive soft particles~\cite{ohern2003,liu_review,van_hecke_review}, variations of the characteristic frequency of disorder-induced vibrational modes, conventionally denoted by $\omega^*$, can be universally quantified in terms of the pressure to bulk elastic modulus ratio \cite{matthieu_thesis, stefanz_pre}. In Fig.~\ref{kappa_vs_omega_fig}c we show that this ratio remains remarkably constant throughout the entire parent temperature range, while the characteristic energy $\kappa$ varies by nearly a factor of $2$. The apparent decoupling of these two observables raises questions about possible essential differences between QLE and the soft modes that emerge near the unjamming point. Understanding these differences is an interesting topic left for future investigation.

\begin{figure}[!ht]
\centering
\includegraphics[width = 0.50\textwidth]{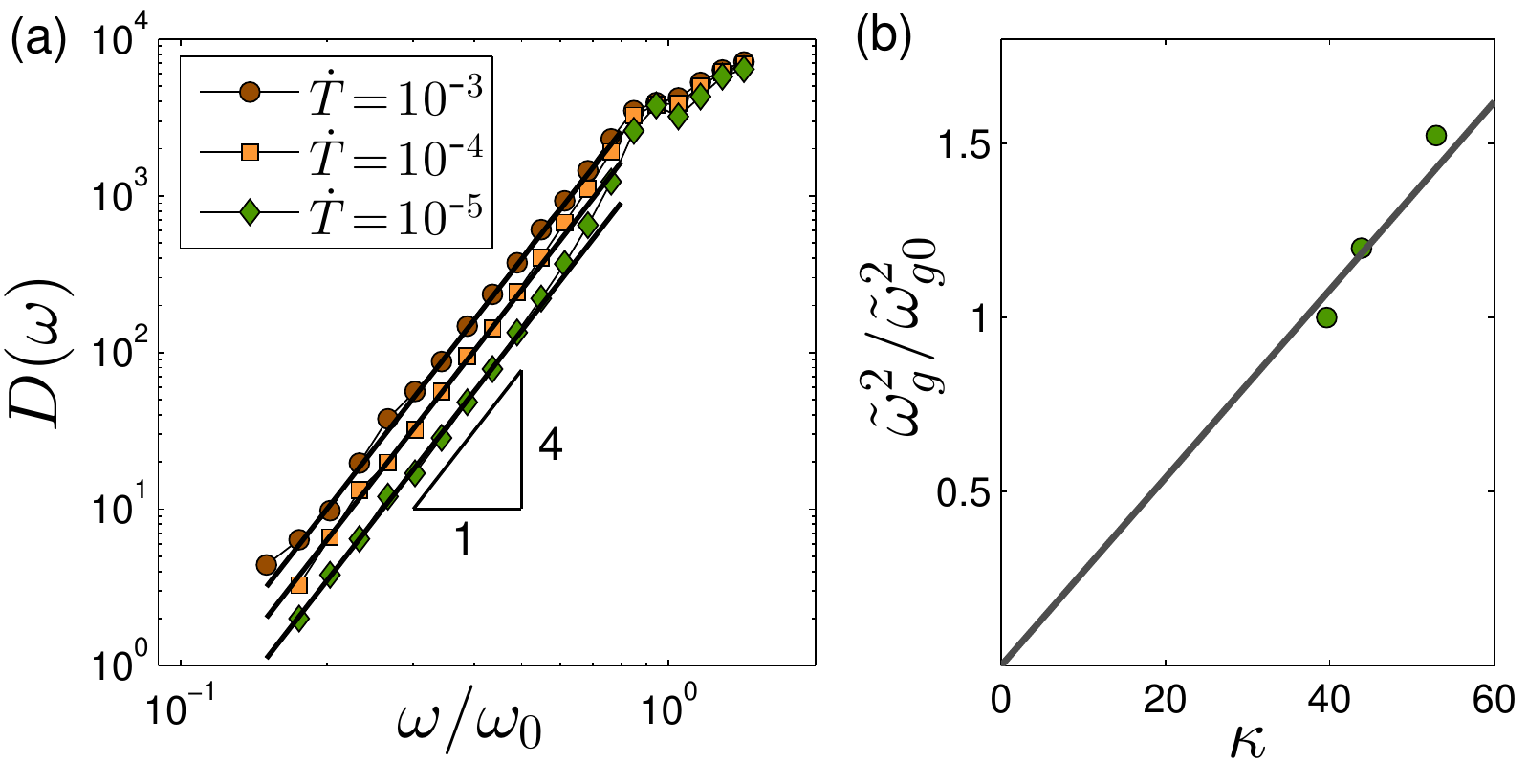}
\caption{\footnotesize (a) Low-frequency tails of the density of vibrational modes $D(\omega)$, measured in 3D glassy samples cooled at rates $\dot{T}\!=\!10^{-3}$ (circles), $\dot{T}\!=\!10^{-4}$ (squares), and $\dot{T}\!=\!10^{-5}$ (diamonds). Using these distributions we extract an energy scale $\tilde{\omega}_g^2$ from the prefactors $\tilde{\omega}_g^{-5}$ of the $\omega^4$ scaling law, see lengthy discussion about the nature of these prefactors in the text. The frequency axis is rescaled by $\omega_0\!=\!2.0$ for visualization purposes. In panel (b), we plot $\tilde{\omega}_g^2$ (normalized by $\tilde{\omega}_{g0}^2\!\equiv\!\tilde{\omega}_{g}^2(\dot{T}\!=\!10^{-3}$)) vs.~the CGEs $\kappa$ measured in each ensemble. Here, consistent with our findings for the minimal vibrational frequencies plotted in Fig.~\ref{kappa_vs_omega_fig}a, we find a very similar relative variation of $\tilde{\omega}_g^2$ and $\kappa$. The continuous line is a guide to the eye.}
\label{continuous_quench_fig}
\end{figure}

We have also investigated the behavior of the CGE $\kappa$ in ensembles of continuously-quenched 3D glassy samples. The low-frequency tails of density of vibrational modes of these ensembles features the universal $\omega^4$ law, as shown in Fig.~\ref{continuous_quench_fig}a. This stands in contrast to the ensembles of instantaneously-quenched glassy samples, that could instead follow $D(\omega)\!\sim\!\omega^\chi$ with $\chi\!<\!4$, as we recently showed in~\cite{protocol_prerc}. This universality allows one to naively extract an energy scale from the prefactor of their respective spectra, which, to this aim, are expressed as $D(\omega)\!\propto\!\tilde{\omega}_g^{-5}\omega^4$.

As discussed in detail in Sect.~\ref{characteristic_glassy_energy}, however, the energy scale $\tilde{\omega}_g^2$ obtained in this way is affected by both the variations in the characteristic energy of QLE, and by their protocol-dependent depletion. Despite that these two effects cannot be straightforwardly disentangled, we superimpose fits to the $\omega^4$ regime in Fig.~\ref{continuous_quench_fig}a, and convert the fitted prefactors into energies $\tilde{\omega}_g^2$, which are in turn plotted in Fig.~\ref{continuous_quench_fig}b against the CGEs $\kappa$ measured in the same ensembles. The data are consistent with our findings for the minimal vibrational frequencies and their relation to the CGE, as shown in Fig.~\ref{kappa_vs_omega_fig}a. Also here, we find that while the two observables are not perfectly proportional to each other, they display very similar relative variations with the quench rate.

Taken together, the results presented in this section for both instantaneously- and continuously-quenched glasses strongly indicate the existence of a relation between the CGE and the softest quasilocalized excitation in the glass, either through the dependence on the parent temperature $T_0$ in the former or on the quench rate $\dot{T}$ 
in the latter. In the next section we firmly establish this connection by considering the system size dependence of the CGE and the energies of the softest QLE.

\section{Establishing a connection between the characteristic energy scale and the softest quasilocalized excitation}
\label{sec:connection}

In the previous section we provided evidence for the existence of an intrinsic connection between the CGE and the energies of the softest glassy QLE by comparing the dependence of these quantities on the degree of annealing of glasses. In order to substantiate this connection, we shift our focus here to their system size dependence. The system size dependence of the softest glassy QLE may emerge from two independent sources. First, there exists an extreme-value statistics consideration which predicts that the energy of the softest (i.e.~minimal) QLE decreases with increasing system size (see quantitative prediction below). In addition to this ``statistical effect", there might also exist a ``mechanical effect'' in which the long-ranged elastic fields associated with QLE, cf.~Fig.~\ref{introduction_fig}, give rise to an extra system size dependence.

As stressed above, the existence, properties and statistics of QLE are independent of whether they can be realized as harmonic modes or not. To highlight this basic idea, which will also have practical implications in the analysis below, we denote by $e$ the energy of QLE. The energy $\omega^2$ of harmonic modes is then generalized to represent the energy of QLE through the relation $e\=\omega^2$. The latter relation can be readily used to obtain the statistical properties of $e$ in light of the universal law $D(\omega)\!\sim\!\omega^4$, which implies $D(e)\!\sim\!e^{3/2}$. With this result at hand, we can predict the system size dependence of the softest glassy QLE associated with the statistical effect. This is achieved through a conventional (Weibullian) extreme-value statistics argument that predicts~\cite{modes_prl}
\begin{equation}
 \label{weibull}
 \int_0^{e_{\mbox{\tiny min}}}e^{3/2}\,de \sim N^{-1} \quad \Rightarrow \quad e_{\mbox{\tiny min}}(L) \sim L^{-2\dbar/5}\ ,
\end{equation}
where $e_{\mbox{\tiny min}}(L)$ is expected to properly represent the $L$-dependence of $e(L)$.

In order to distinguish between the statistical and mechanical effects on the $L$-dependence of $e$, our goal now is to calculate $e_{\mbox{\tiny min}}$ and test it against Eq.~\eqref{weibull}. If the prediction is verified, we know there exists no significant mechanical contribution to the $L$-dependence; otherwise, it exists. We need to use different procedures to calculate $e_{\mbox{\tiny min}}$ in 2D and 3D. In 3D, for systems that are small enough to suppress the occurrence of phonons at very low frequencies, the softest QLE assume the form of \emph{harmonic} vibrational modes, as demonstrated in~\cite{modes_prl} and further discussed in Appendix~\ref{sec:2D}. For this reason, in the previous section we could identify the energies of the softest QLE with those of the lowest-frequency vibrational modes. Quantitatively extracting the energies of the softest QLE from harmonic analyses in 2D is less straightforward. In Appendix~\ref{sec:2D} we show that many harmonic vibrational modes associated with minimal vibrational frequencies are in fact poor representatives of QLE in 2D glasses due to mixing with phonons of similar frequencies. Consequently, as soft QLE are not well-represented by vibrational harmonic modes in 2D we resort to nonlinear measures which offer a robust definition for the softest QLE.

Therefore, to quantify $e_{\mbox{\tiny min}}$ in 2D, we consider a family of {\em nonlinear} quasilocalized excitations, as put forward in~\cite{SciPost2016}, and referred to in what follows as quartic modes. The latter are solutions $\hat{\pi}_4$ to the nonlinear equation
\begin{equation}
{\cal M}\cdot\hat{\pi}_4 = \frac{{\cal M}\!:\!\hat{\pi}_4\hat{\pi}_4}{\frac{\partial^4U}{\partial\vec{x}\partial\vec{x}\partial\vec{x}\partial\vec{x}}\!::\!\hat{\pi}_4\hat{\pi}_4\hat{\pi}_4\hat{\pi}_4}\frac{\partial^4U}{\partial\vec{x}\partial\vec{x}\partial\vec{x}\partial\vec{x}}\tripleCdot\hat{\pi}_4\hat{\pi}_4\hat{\pi}_4\,,
\end{equation}
where the symbols $:,\tripleCdot$ and $::$ denote double, triple and quadruple contractions, respectively. In Appendix~\ref{definitions} we elaborate more about our calculations of quartic modes in glassy samples. The key advantage of considering quartic modes in the present context is their absolute indifference to the presence of phonons of similar energies, as shown in e.g.~\cite{SciPost2016}. This allows us to access one of the lowest energy QLE in 2D systems and hence to quantitatively determine its value, regardless of the proximity of its energy to that of phonons. Specifically, $e_{\mbox{\tiny min}}$ is calculated through ${\cal M}\!:\!\hat{\pi}_4\hat{\pi}_4$.

\begin{figure}[!ht]
\centering
\includegraphics[width = 0.50\textwidth]{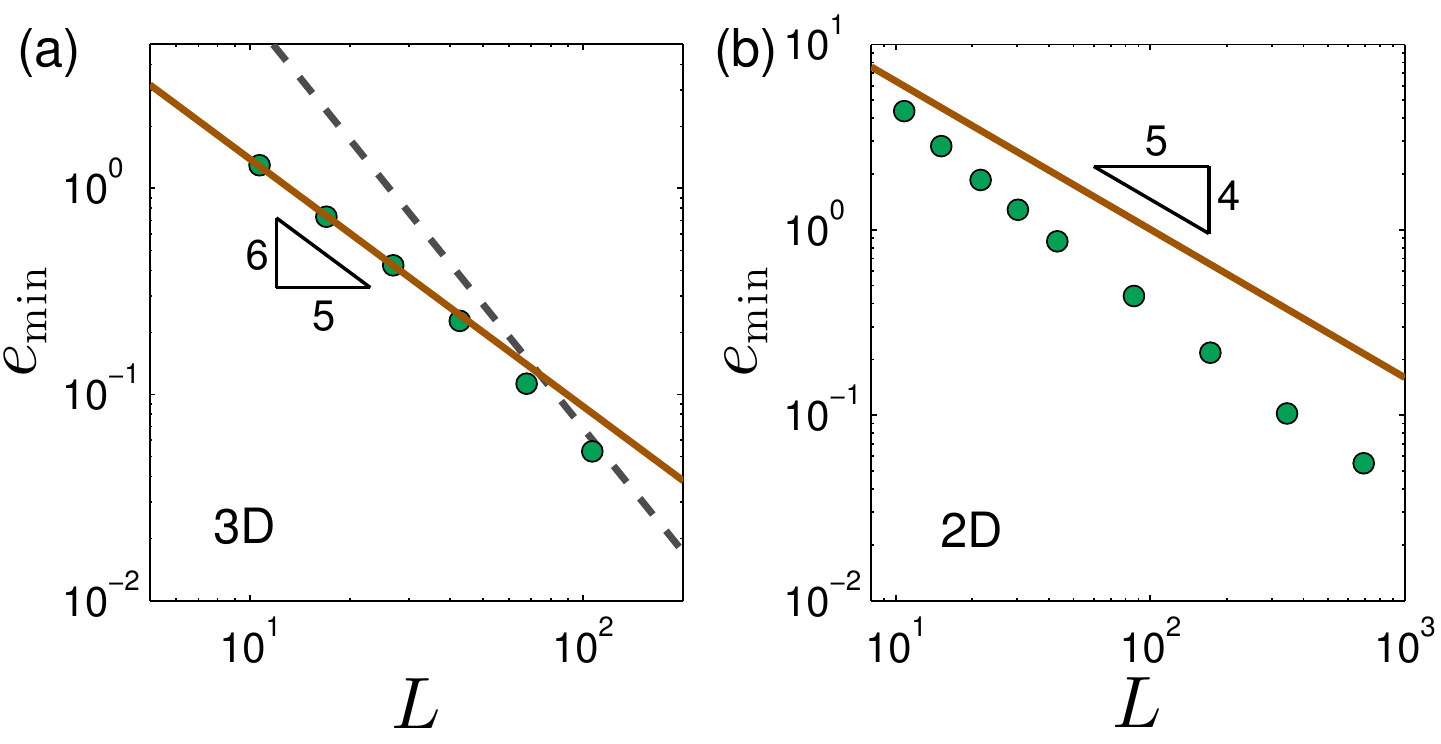}
\caption{\footnotesize Sample-to-sample means of the energy $e_{\mbox{\tiny min}}$ of the softest QLE vs.~system size $L$, for glassy samples in (a) 3D and (b) 2D. For each 3D glassy sample $e_{\mbox{\tiny min}}$ is identified with the energy of the lowest-frequency harmonic vibrational mode $\omega_{\mbox{\tiny min}}^2$. We note that $\omega_{\mbox{\tiny min}}^2$ is bounded from above by the energy of the softest phonon $4\pi^2G/(\rho L^2)$, represented in panel (a) by the dashed line. In 2D, $e_{\mbox{\tiny min}}$ represents the energy of a soft quartic mode, calculated as described in the text. The dependence of $e_{\mbox{\tiny min}}$ on system size follows the Weibullian expectation $e_{\mbox{\tiny min}}\!\sim\!L^{-2\dbar/5}$ in 3D, but not in 2D.}
\label{only_min_mode_fig}
\end{figure}

In Fig.~\ref{only_min_mode_fig} we test the prediction on the right part of Eq.~\eqref{weibull}, where the harmonic frequencies $\omega_{\mbox{\tiny min}}^2$ are used to calculate $e_{\mbox{\tiny min}}$ in 3D (left panel) and   ${\cal M}\!:\!\hat{\pi}_4\hat{\pi}_4$ is used for that purpose in 2D (right panel). It is observed that Eq.~\eqref{weibull} is approximately followed in 3D ($\dbar\=3$), but seriously fails in 2D ($\dbar\=2$). These results indicate that the mechanical effect is rather weak in 3D, because the statistical considerations alone rather accurately predict the system size dependence in 3D, while it is significantly stronger in 2D. Consequently, we focus on 2D in the remainder of this section.

\begin{figure}[!ht]
\centering
\includegraphics[width = 0.50\textwidth]{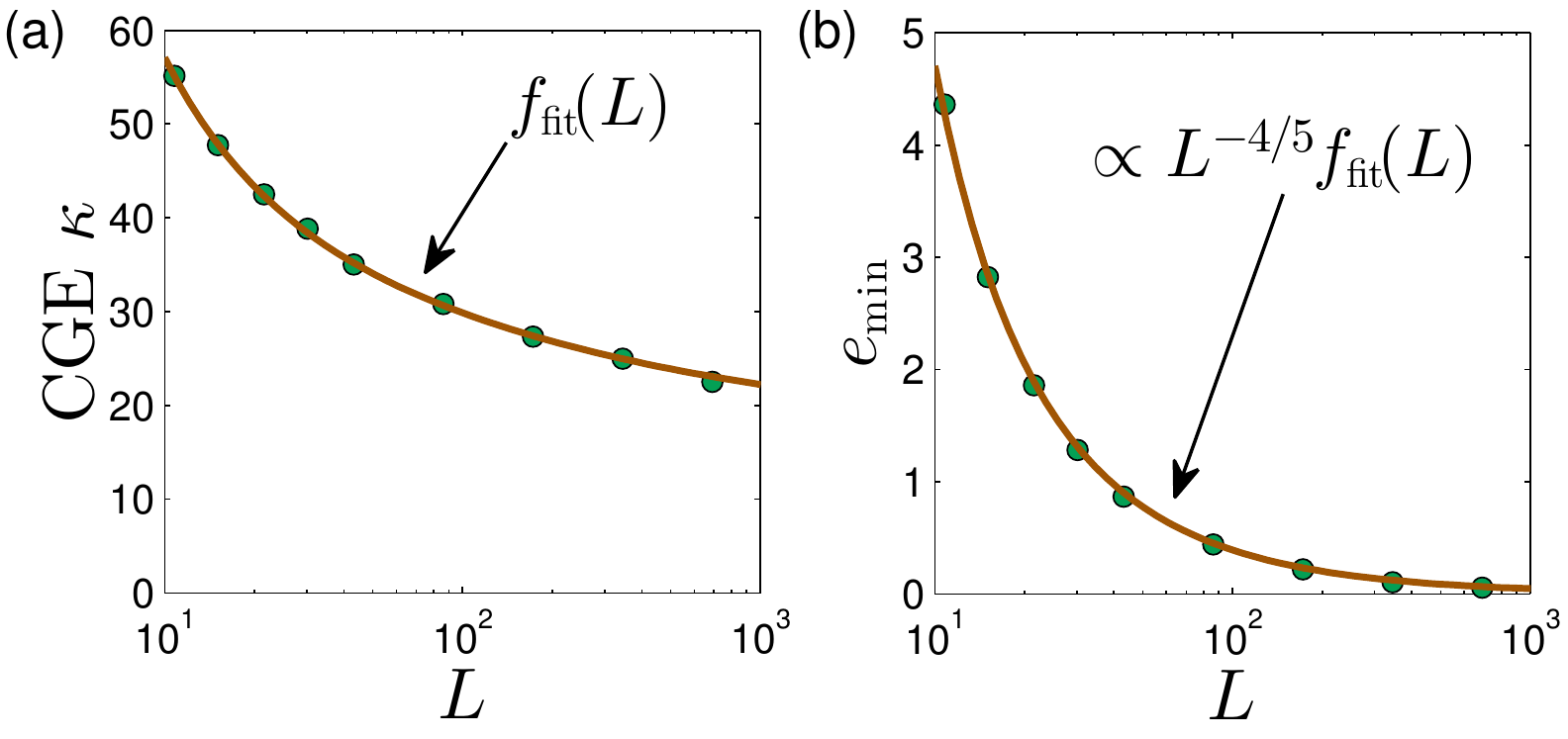}
\caption{\footnotesize (a) The characteristic energy $\kappa$ measured in 2D glassy samples. We superimposed in this plot a fitting function $f_{\mbox{\tiny fit}}(L)$, see text for details. (b) Sample-to-sample mean energies of soft quartic modes. Notice the order-of-magnitude difference between the scales of the $y$-axis of the two panels. The agreement of the product $L^{-4/5}f_{\mbox{\tiny fit}}(L)$ with our data strongly supports our suggested combination of statistical and mechanical effects on quartic modes' energies, and puts our proposition that QLE energies are well-captured by $\kappa$ on firm grounds.}
\label{QLE_vs_dipole_fig}
\end{figure}

The mechanical effect on the $L$-dependence of the QLE energy $e(L)$ in 2D is theoretically expected to be fully contained within the $L$-dependence of the CGE $\kappa(L)$, which should be completely independent of the statistical effect quantified in Eq.~\eqref{weibull}. Figure~\ref{QLE_vs_dipole_fig}a presents the means of $\kappa(L)$, where we superimpose on the data a fitting function $f_{\mbox{\tiny fit}}(L)$~\footnote[2]{We use $f_{\mbox{\tiny fit}}(L)\!=\! c_1L^{-e_1}\!+\! c_2L^{-e_2} \!+\! c_3$, with the following fitting parameters: $c_1\!=\!59.64$, $c_2\!=\!354.1$, $c_3\!=\!15.46$, $e_1\!=\!0.3156$, and $e_2\!=\!1.442$}; the function $f_{\mbox{\tiny fit}}(L)$ will be theoretically derived in the next section, but its precise form and physical origin are not crucial for our discussion here. The important point to note is that $\kappa(L)$ exhibits a significant variation with the system size $L$ and consequently we predict that
\begin{equation}
\label{eq:combined}
e_{\mbox{\tiny min}}(L) \sim L^{-4/5}\kappa(L)\,,
\end{equation}
i.e.~$e_{\mbox{\tiny min}}(L)$ is predicted to be a product of the statistical $L$-dependence predicted in Eq.~\eqref{weibull} and of the mechanical $L$-dependence embodied in $\kappa(L)$.

The prediction in Eq.~\eqref{eq:combined} is tested in Fig.~\ref{QLE_vs_dipole_fig}b by superimposing the function $L^{-4/5}f_{\mbox{\tiny fit}}(L)$ on the data for $e_{\mbox{\tiny min}}(L)$, demonstrating excellent agreement. This {\em quantitative} agreement between the combined statistical and mechanical effects on soft QLE further --- and strongly --- reinforces our key proposition that the scale defined by the response to local force dipoles intrinsically captures the characteristic scale of soft QLE. Together with our demonstration of such a relation for annealed glasses in the previous section, this result establishes a nontrivial relation between rare fluctuations --- the soft quartic modes' energies ---, and a bulk response --- the characteristic energies. In the next section we theoretically predict the $L$-dependence of the mechanical effect embodied in $\kappa(L)$ along with its physical origin, as part of a broader discussion of the spatial dimension and system size dependence of the characteristic energy scale.

\section{Long-ranged elastic effects of quasilocalization: System size and spatial dimensionality dependence of the characteristic energy scale}
\label{system_size_section}

The discussion of the system size dependence of the CGE $\kappa(L)$ in 2D allowed us to establish a remarkable relation between the CGE and the softest QLE in the previous section. The dependence itself, expressed through a fitting function $f_{\mbox{\tiny fit}}(L)$, and its physical origin have not been addressed. The dipolar response, quantified by the CGE, and QLE --- shown in Fig.~\ref{introduction_fig} --- are quasilocalized, i.e.~they are characterized by a disordered core with a characteristic lengthscale of a few atomic sizes and in addition feature long-ranged elastic fields. The observed $L$-dependence of $\kappa(L)$ is expected to emerge from the latter. Hence, our first goal in this section is to theoretically derive this $L$-dependence in 2D and demonstrate its direct connection to the long-ranged elastic fields associated with quasilocalization.

To this aim, we start by considering Eq.~\eqref{eq:dipolar_response} whose solution $\vec{z}$ is the response to a force dipole $\vec{d}$. At the continuum level, which is relevant to the long-ranged contribution to $\vec{z}$, the response to a localized force is given by the elastic Green's function ${\bm G}({\bm r})$~\cite{landau1986theory}. Consequently, the response to a force dipole is determined by the spatial derivative of ${\bm G}({\bm r})$, $\vec{z}\!\sim\!\partial_r {\bm G}({\bm r})\!\sim\!r^{1-\dbar}$, where $r$ is the distance from the core. To use this scaling relation in the expression for $\kappa_\alpha$ in Eq.~\eqref{foo00}, we first note that its numerator, $\vec{z}\cdot{\cal M}\cdot\vec{z}\=\vec{z}\cdot\vec{d}$, is a local contribution that is independent of the system size $L$ due to the contraction with the dipole. Consequently, the $L$-dependence of $\kappa$ in Eq.~\eqref{foo00} is determined by the denominator, that is
\begin{equation}
\label{eq:Lscaling}
\frac{1}{\kappa(L)} \sim \vec{z}\cdot\vec{z} \sim \int_{\xi_\kappa}^L r^{2(1-\dbar)} r^{\dbar-1} dr \sim \int_{\xi_\kappa}^L r^{1-\dbar} dr \ ,
\end{equation}
where $\xi_\kappa$ is an atomic-scale cutoff length characterizing the disordered core and $r^{\dbar-1} dr$ is a volume element in $\dbar$ dimensions. Note that we used the infinite system Green's function ${\bm G}({\bm r})$ and that the finite system size effect emerges from the upper integration limit.

\begin{figure}[!ht]
\centering
\includegraphics[width = 0.50\textwidth]{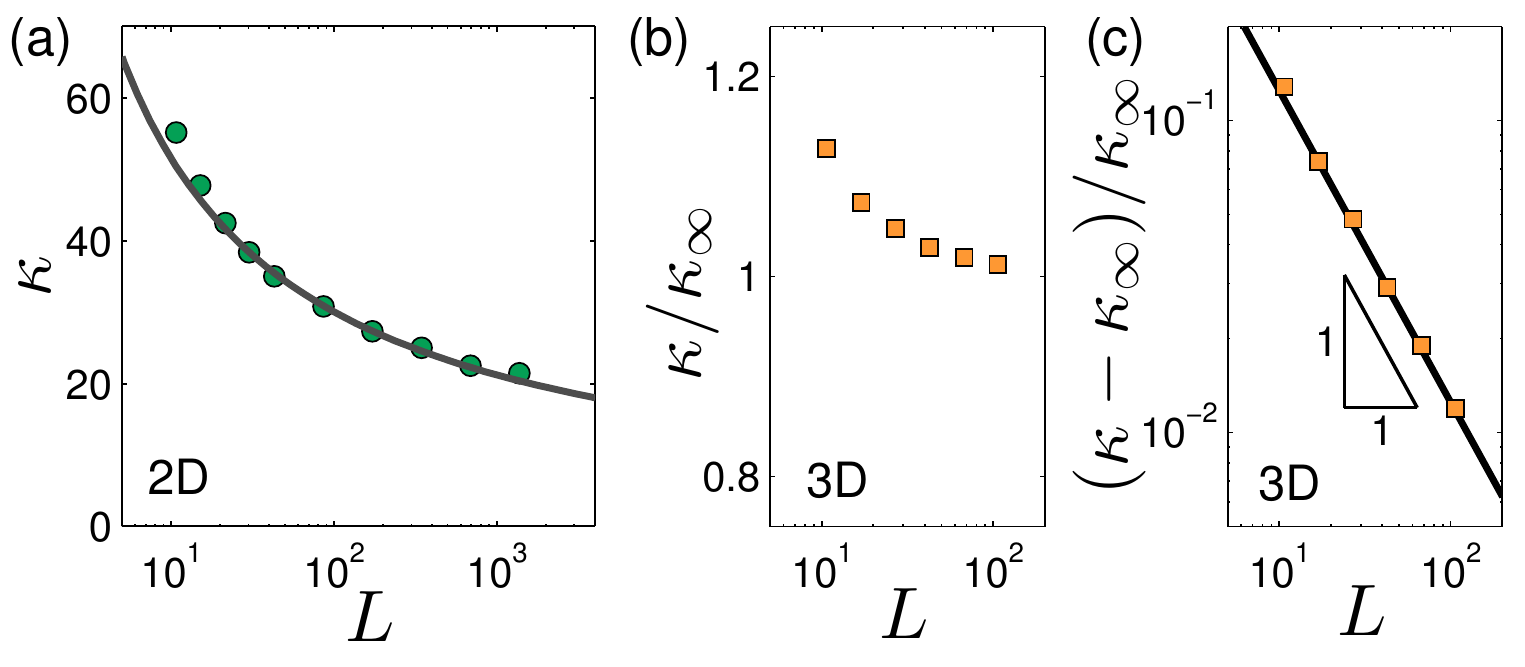}
\caption{\footnotesize The symbols show characteristic energies measured in our computer glass model in 2D (a) (same as Fig.~\ref{QLE_vs_dipole_fig}a) and in 3D (b),(c). The continuous lines in panels (a) and (c) follow the theoretical predictions given by Eqs.~(\ref{foo03}) and (\ref{foo04}), respectively.}
\label{dipole_stiffness_N_dependence}
\end{figure}

Applying Eq.~\eqref{eq:Lscaling} in 2D ($\dbar\=2$), we obtain
\begin{equation}
\label{foo03}
\kappa(L) \sim \frac{1}{\log (L/\xi_\kappa)}\,, \quad \mbox{in 2D} \ .
\end{equation}
This prediction is shown in Fig.~\ref{dipole_stiffness_N_dependence}a to be in excellent quantitative agreement with the numerical data for $\kappa(L)$ in 2D (the same data were presented in Fig.~\ref{QLE_vs_dipole_fig}a), demonstrating that one can analytically predict the previously used fitting function $f_{\mbox{\tiny fit}}(L)$. Equation~\eqref{foo03} indicates that in 2D solids the stiffness associated with the response to a local force dipole \emph{vanishes} in the thermodynamic limit $L\!\to\!\infty$, albeit very slowly (logarithmically). This peculiarity might be related to other 2D peculiarities in glass physics, as will be briefly mentioned below.

Equation~\eqref{eq:Lscaling} can also be used to elucidate the dimensionality dependence of the CGE. To this aim, we apply it in 3D ($\dbar\=3$) to obtain the prediction
\begin{equation}
\label{foo04}
\kappa(L) - \kappa(L\!\to\!\infty) \sim L^{-1}\,, \quad \mbox{in 3D} \ ,
\end{equation}
which is quantitatively verified in Figs.~\ref{dipole_stiffness_N_dependence}b,c. Equation~\eqref{foo04} shows that $\kappa$ approaches a {\em finite value} in the thermodynamic limit $L\!\to\!\infty$. Moreover, the small variation of $\kappa(L)$ observed in Figs.~\ref{dipole_stiffness_N_dependence}b,c explains why the statistical $L$-dependence dominates the mechanical one in 3D, as has been already demonstrated in Fig.~\ref{only_min_mode_fig}a. The origin of the significant spatial dimensionality dependence of the CGE $\kappa(L)$, and most notably the anomalous $1/\log(L)$ behavior in 2D, is the long-ranged elastic fields that decorate the localized disordered core of QLE. This peculiarity appears to be intimately related to the 2D elastic Green's function ${\bm G}(r)\!\sim\!\log(r)$, and as such is reminiscent of the recent discussion of long-wavelength Mermin-Wagner-like fluctuations in 2D glasses and their relation to possible differences between the glass transition in 2D and 3D~\cite{Szamel2015,Keim_2017,dimension_dependence_of_glass_transition_chaikin_pnas_2017,Tarjus_commentary}. This potentially interesting connection should be further explored in the future.

\section{The characteristic energy scale and glassy slowing down}
\label{sec:relaxation}

Having established the existence of an intimate relation between the characteristic energy scale $\kappa$ and the energy scale of the softest quasilocalized excitations, and that both energies show large variations with annealing, we next build on these observations and test possible relations between the CGE and relaxation times~of the supercooled liquid. Previous work~\cite{Schober_correlate_modes_dynamics, wyart_brito_2007, widmer2008irreversible, harrowell_2009} has shown that there are strong spatial correlations between quasilocalized soft vibrational modes and regions in the supercooled liquid that are more susceptible to thermally-driven rearrangements. It is therefore plausible to expect that the increase in the energies of QLE upon supercooling slows down structural relaxation processes.

\begin{figure}[!ht]
\centering
\includegraphics[width = 0.50\textwidth]{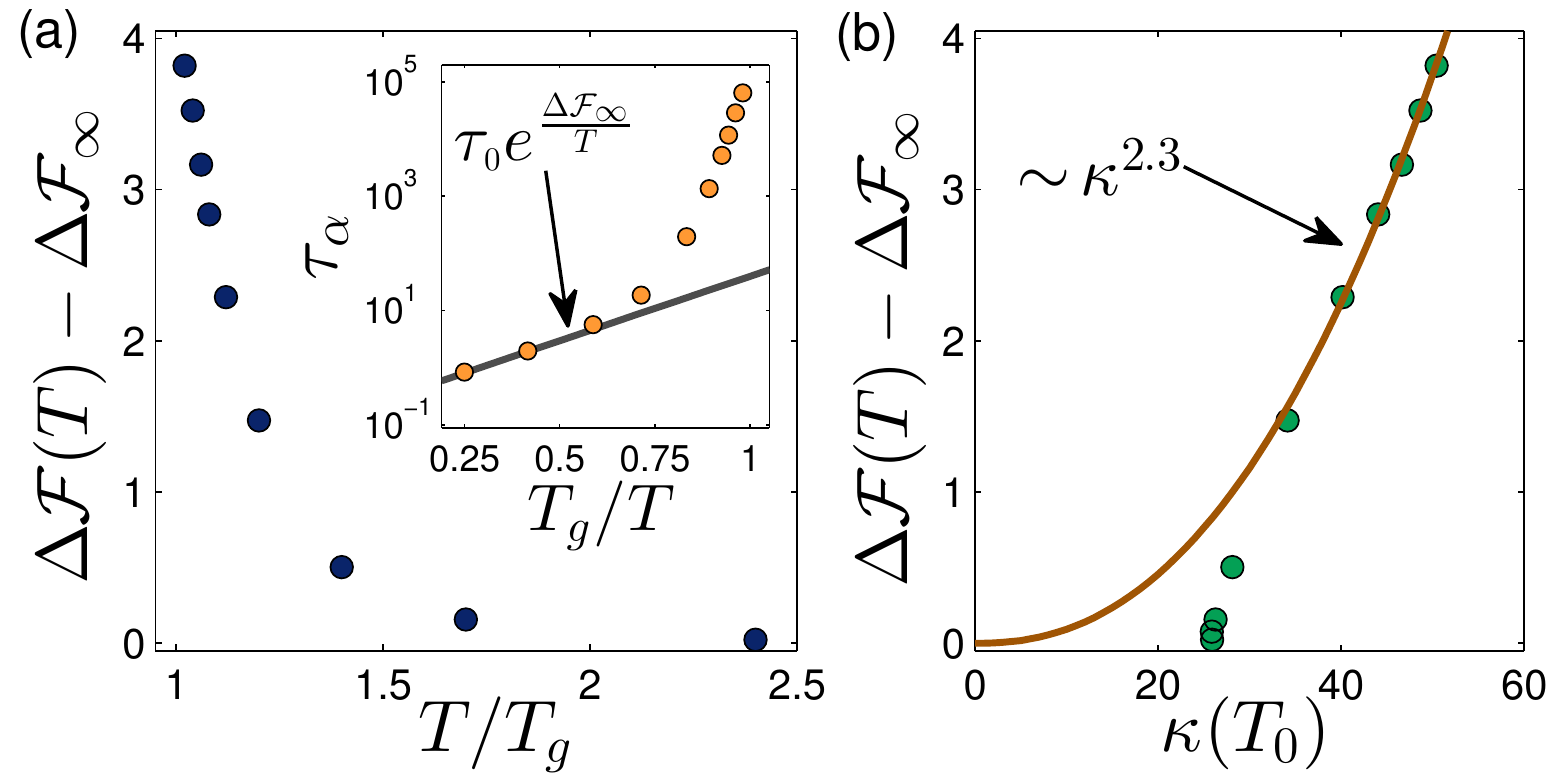}
\caption{\footnotesize (a) Excess activation barriers $\Delta{\cal F}\!-\!\Delta{\cal F}_\infty$ vs.~temperature. $\Delta{\cal F}_\infty$ is read off the Arrhenius fit of the relaxation times $\tau_\alpha\!=\!\tau_0\exp(\Delta{\cal F}_\infty/T)$, as shown in the inset, and $\Delta {\cal F}\!\equiv\!T\log(\tau_\alpha/\tau_0)$. See text for the fitted values of $\tau_0$ and $\Delta{\cal F}_\infty$. (b) Excess activation barriers vs.~dipole energies~$\kappa$, measured in the underlying inherent states of the equilibrium liquid at each temperature.}
\label{relaxation_barrier_kappa_relation_fig}
\end{figure}

In Fig.~\ref{relaxation_barrier_kappa_relation_fig}a we plot the excess activation barriers $\Delta{\cal F}$ over the Arrhenius part $\Delta{\cal F}_\infty$, extracted from the structural relaxation times $\tau_\alpha$ (see inset and also Fig.~\ref{alpha_relaxation_fig}), vs.~temperature $T$. The fitted values that appear in the Arrhenius form $\tau_\alpha\!=\!\tau_0\exp(\Delta{\cal F}_\infty/T)$ are $\tau_0\!=\!0.22$ and $\Delta{\cal F}_\infty\!=\!2.6$. In Fig.~\ref{relaxation_barrier_kappa_relation_fig}b we plot the same excess activation barriers as shown in panel (a), but this time against the CGE $\kappa(T_0)$ measured in the underlying inherent states of the corresponding equilibrium temperatures.

It is expected that the dynamics of the high-temperature liquid are indifferent to the value of the characteristic energies measured in underlying inherent states. In fact, we only expect to find some connection between properties of underlying inherent states and the dynamics of the supercooled liquids below the onset temperature of the so-called two-step relaxation and the deviation from Arrhenius-like relaxation, which occurs in our system at around $T\!\approx\!0.70$ for which $\kappa\!\approx\!28$. We superimpose on the data of Fig.~\ref{relaxation_barrier_kappa_relation_fig}b a fit to the scaling $\kappa^{2.3}$. While other functional forms may provide better fits to a larger range of temperatures, for reasons motivated below we restrict the fitted function to a simple form that vanishes with vanishing $\kappa$.

How can the observed relation between the characteristic energy scale and activation barriers be understood? Here we propose a very simple scenario, which draws inspiration from the Soft Potential Model \cite{soft_potential_model_1991,soft_potential_model_1987,chalker}, and elastic models for relaxation in supercooled liquids \cite{elastic_model1996,elastic_model2004,elastic_model2006,jeppe_review2006}. We consider a QLE with an associated energy $\kappa$, and expand the potential energy to third order in terms of the displacement $s$ along the QLE as
\begin{equation}
U \simeq U_0 + \sFrac{1}{2}\kappa s^2 + \sFrac{1}{6}\tau s^3 + {\cal O}(s^4)\,,
\end{equation}
where $\tau\!\equiv\!\frac{\partial^3U}{\partial\vec{x}\partial\vec{x}\partial\vec{x}}\!\tripleCdot\!\hat{z}\hat{z}\hat{z}$, and $\hat{z}$ is the QLE field. Within this approximation, the potential barrier $\Delta U$ that separates neighboring inherent states is expressed in terms of the coefficients $\kappa$ and $\tau$ as
\begin{equation}\label{naive}
\Delta U \simeq \frac{2}{3}\frac{\kappa^3}{\tau^2}\,.
\end{equation}

Naively, this simple scenario suggests that activation barriers could grow as a relatively large ($\approx\!3$) power of the energies $\kappa$, as we indeed find. However, as mentioned above in the context of our discussion about glassy lengthscales, the core-size of QLE decreases with deeper supercooling \cite{protocol_prerc}; one then expects the third order coefficient $\tau$ to also vary with supercooling. In particular, they are expected to grow with stronger localization of QLE, as shown in \cite{plastic_modes_prerc, SciPost2016}, which implies they would increase with increasing $\kappa$. Interestingly, in the Soft Potential Model \cite{soft_potential_model_1991,soft_potential_model_1987,chalker} it is argued that stability requires that the third order coefficients $\tau$ are bounded in magnitude by $\sqrt{\kappa}$. Adopting this simple scenario, this would suggest that $\Delta U\!\sim\!\kappa^2$, in reasonable agreement with the data. These results suggest the existence of a direct connection between the CGE and the slowing down of relaxational dynamics with increased supercooling. This fundamental relation should be systematically explored in the future using extensive computer simulations.

\section{discussion}
\label{discussion}

In this work we put forward and tested the proposition that the response to a local force dipole defines a characteristic energy scale that plays a key role in determining several important structural and dynamical features of glasses and deeply supercooled liquids. Here we discuss our findings in the context of other current work on glasses and supercooled liquids, and suggest future research directions.

\subsection{Annealing dependence}
In our work we showed that the characteristic energy scale changes dramatically in inherent states that underlie equilibrium supercooled configurations as the glass transition temperature is approached: its relative variation is about 4 times larger compared to the variation of elastic moduli in the same glassy samples. These observations strongly support the assertion by Wyart~\cite{wyart_vibrational_entropy} that annealing not only leads to the stiffening of elastic moduli, but is also accompanied by the increase of other characteristic energy scales, in particular of the boson peak frequency. In the same work Wyart put forward a relation between the fragility of a glass former and accompanying inhomogeneity of relative variations of typical frequency scales upon annealing, with larger inhomogeneities expected in more fragile systems. It would be interesting to test this relation computationally using the ideas introduced in this work, and in particular to verify whether the characteristic glassy energy scale as defined by the response to local force dipoles displays larger relative variations with annealing in more fragile glass forming models.

\subsection{Relation to soft QLE and supercooled liquids' dynamics}

We further established a nontrivial relation between the characteristic glassy energy scale, and energies of the softest quasilocalized vibrational modes observed in samples made with different degrees of annealing. We showed that despite possessing vastly different scales, separated by roughly two orders of magnitude for the system size employed, these two energies exhibit similar relative variations with annealing. The separation between these scales is $L$-dependent, and would further increase by increasing $L$. Following these observations, and further motivated by previous work~\cite{Schober_correlate_modes_dynamics, wyart_brito_2007, widmer2008irreversible,harrowell_2009} that established spatial and geometric correlations between soft, quasilocalization excitations and structural relaxation in supercooled liquids, we examined the relation between the characteristic energy scale and activation barriers towards structural relaxation.  We found that below the crossover temperature to the `landscape dominated' regime~\cite{landscape_dominated_jeppe_2000}, activation barriers as extracted from dynamics grow roughly as $\kappa^{2.3}$, where the characteristic energy scale $\kappa$ was measured in the underlying inherent states of equilibrium configurations. This result is reasonably consistent with simple scaling arguments inspired by the Soft Potential Model framework \cite{soft_potential_model_1991,soft_potential_model_1987,chalker}. The refinement of these arguments is left for future research efforts.

The comparison between energies of the softest QLE and the characteristic glassy energy scale (see Fig.~\ref{kappa_vs_omega_fig}) suggests that in addition to the increase of these scales upon supercooling, the density of QLE is also depleted. In studying the possible relation between characteristic glassy scales and activation barriers, we did not account for entropic effects that could stem from the depletion of QLE, i.e.~from the changes in the number of accessible relaxation directions in the PEL, upon deep supercooling. Resolving the precise relative roles played by the depletion of QLE vs.~the increase in their characteristic energies, in determining activation barriers towards structural relaxation, is an important topic proposed for future research.

Our examination of a possible relation between the characteristic glassy scale and activation barriers is inspired by similar ideas put forward within the framework of elastic models of supercooled liquids by Dyre and coworkers \cite{elastic_model1996,elastic_model2004,elastic_model2006,jeppe_review2006}. The viewpoint adopted in these models is that the dominant contribution to activation barriers is some form of \emph{elastic} energy. Further support for this viewpoint was recently suggested in \cite{Wyart_Cates_prl_2017,swap_arXiv_2018}. To motivate this viewpoint in the context of elastic models, it is highlighted that relaxation events predominantly involve shear deformation, and therefore the relevant energy scale of activation barriers is set by the (high-frequency) shear modulus. The dominance of elastic energy in determining activation barriers is tested in our study as well; however, differently from elastic models, we speculate that barriers towards relaxation are not controlled by macroscopic elasticity (at least not in 3D, see further discussion below), but instead by the intrinsically-atomistic, micromechanical elastic properties of quasilocalized excitations, which, in turn, vary substantially under annealing.

We further note that a related viewpoint that attributes a micromechanical, elastic origin to the dramatic slowing down of dynamics in dense colloidal glasses has been proposed by Brito and Wyart \cite{wyart_brito_2007,wyart_brito_2009}. These authors argue that the onset of activated dynamics in colloidal glasses stems from the stabilization of the softest collective degrees of freedom, which were observed to feature an extended spatial character. Here we do not study the onset of activated dynamics, but instead test the assumption that the properties of local minima of the potential energy landscape that underlie equilibrium states are relevant for activated relaxation at temperatures lower than the crossover temperature to the landscape-dominated regime. Understanding the extent to which this assumption holds, and the intricate details of the transition to landscape-dominated activated dynamics is an important goal for future research.

\subsection{Relation to the unjamming scenario}

It was previously proposed in \cite{new_variational_argument_epl_2016}, and tested in \cite{new_variational_argument_epl_2016,stefanz_pre}, that the energy scale that characterizes soft, extended vibrational modes that emerge near the unjamming transition --- the loss of solidity observed in soft-sphere packings as the confining pressure is reduced \cite{ohern2003,liu_review,van_hecke_review} --- can be defined by the response to local force dipoles. The statistical and structural properties of soft, extended vibrational modes that emerge near the unjamming point are predicted by mean-field frameworks \cite{wyart_emt,eric_boson_peak_emt,eric_hard_spheres_emt,mean_field_glass_theory_prl_2016_kurchan_zamponi, mean_field_glass_theory_j_stat_mech_2016_kurchan_zamponi, mean_field_glass_theory_pre_2016_Zamponi,silvio} and variational arguments \cite{eric_hard_spheres_emt,new_variational_argument_epl_2016}; intriguingly, they are qualitatively and quantitatively different from the statistical and structural properties of the quasilocalized excitations studied here for a generic structural glass far from the unjamming point~\cite{modes_prl,SciPost2016, protocol_prerc, inst_note}, despite that the same protocol appears to capture the characteristic energies of both types of soft excitations. The essential nature of the relation between these two classes of soft excitations is still an open question that deserves further investigation. Some very recent progress in addressing this question has been made in \cite{ikeda_length_with_MW_2018}. 

\subsection{Spatial dimensionality and finite-size effects}

In addition to the annealing dependence of the characteristic glassy energy scale defined by the response to local force dipoles, we further showed that it exhibits strong spatial dimensionality and system size dependencies: we find that while in 3D no substantial system size dependence is observed, in 2D we observe very strong finite size effects on the characteristic energies, that are expected to vanish in the thermodynamic limit. Similarly to the study of the annealing dependence as discussed above, in our systematic investigation of finite size effects on characteristic energies in 2D we also observe a remarkable relation between the characteristic energy scale --- a bulk average response --- and the energy of soft, quasilocalized excitations, which manifest rare structural fluctuations.

We note importantly that a similar finite-size scaling analysis to that presented in Sect.~\ref{system_size_section} for the $L$-depenedence of the characteristic energies can be carried out for the magnitude of the small potential energy barriers that separate inherent states on the potential energy landscape. A detailed report of this analysis is kept for future work; here we note that our analysis indicates that system size variations in anharmonicities of the potential energy lead to an expected $L$-independence of characteristic potential energy barriers, i.e.~they are not expected to vanish in the thermodynamic limit, despite the apparent vanishing of the characteristic energy scale. We therefore spectulate that the qualitative differences between supercooled liquids' relaxation dynamics in 2D and 3D, as observed in \cite{karmakar_lengthscale2} and recently discussed in \cite{Szamel2015,Keim_2017,dimension_dependence_of_glass_transition_chaikin_pnas_2017, Tarjus_commentary}, does not stem from an explicit $L$-dependence of characteristic potential energy barriers. Instead, we speculate that it rather originates from the different form of the decay of the far fields of quasilocalized excitations.

\acknowledgements
We warmly thank M.~Aldam, G.~D\"uring, E.~DeGiuli and M.~Wyart for fruitful discussions. E.~L.~acknowledges support from the Netherlands Organisation for Scientific Research (NWO) (Vidi grant no.~680-47-554/3259). E.~B.~acknowledges support from the Minerva Foundation with funding from the Federal German Ministry for Education and Research, the William Z.~and Eda Bess Novick Young Scientist Fund and the Harold Perlman Family.

\appendix

\section{Models, protocols, and observables}
\label{appendix}

\subsection{Model glass former and microscopic units}

Most of the computer experiments presented in this work are performed on a generic glass-forming model in 2D and 3D. We employed a 50:50 binary mixture of `large' and `small' particles of equal mass $m$ enclosed in a cubic (square) box in 3D (2D) of linear size $L$. The particles interact via a radially symmetric, purely repulsive pairwise potential of the form
\begin{equation}
\varphi(r_{ij}) = \left\{ \begin{array}{ccc}\varepsilon\left[ \left( \sFrac{\lambda_{ij}}{r_{ij}} \right)^n + \sum\limits_{\ell=0}^q c_{2\ell}\left(\sFrac{r_{ij}}{\lambda_{ij}}\right)^{2\ell}\right]&,&\sFrac{r_{ij}}{\lambda_{ij}}\le x_c\\0&,&\sFrac{r_{ij}}{\lambda_{ij}}> x_c\end{array} \right.,
\end{equation}
where $r_{ij}$ is the distance between the $i^{\mbox{\tiny th}}$ and $j^{\mbox{\tiny th}}$ particles, $\varepsilon$ is a microscopic energy scale, and $x_c$ is the dimensionless distance for which $\varphi$ vanishes continuously up to $q$ derivatives. Distances are measured in terms of the interaction lengthscale $\lambda$ between two `small' particles, and the rest are chosen to be $\lambda_{ij}\!=\!1.18\lambda$ for one `small' and one `large' particle, and $\lambda_{ij}\!=\!1.4\lambda$ for two `large' particles. The coefficients $c_{2\ell}$, determined by demanding that $\varphi$ vanishes continuously up to $q$ derivatives, are given by
\begin{equation}
c_{2\ell} = \frac{(-1)^{\ell+1}}{(2q\!-\!2\ell)!!(2\ell)!!}\frac{(n\!+\!2q)!!}{(n\!-\!2)!!(n\!+\!2\ell)}x_c^{-(n+2\ell)}\,.
\end{equation}
We chose the parameters $x_c\!=\!1.48, n\!=\!10$, and $q\!=\!3$. The density was set to be $N/V\!=\!0.82\lambda^{-3}$ in 3D, and $N/V\!=\!0.86\lambda^{-2}$ in 2D, where $N$ stands for the total number of particles, and $V\!\equiv\!L^\dbar$ is the volume in $\dbar$ spatial dimensions. Time is expressed in terms of $\bar{\tau}\!\equiv\!\sqrt{m\lambda^2/\varepsilon}$, temperature in terms of $\varepsilon/k_B$ with $k_B$ the Boltzmann constant, quench rates in terms of $\varepsilon/(k_B\bar{\tau})$, stress, pressure, and elastic moduli in terms of $\varepsilon/\lambda^{\dbar}$, and vibrational frequencies in terms of~$\bar{\tau}^{-1}$. This model undergoes a glass transition at a temperature of about $T_g\!\approx\!0.5\varepsilon/k_B$, both in 2D and 3D. In the main text and in what follows we omit the units of all observables; these should be understood as expressed in terms of the microscopic units as specified here.

\subsection{Ensembles of glassy samples}
\label{ensembles}

\begin{figure}[!ht]
\centering
\includegraphics[width = 0.45\textwidth]{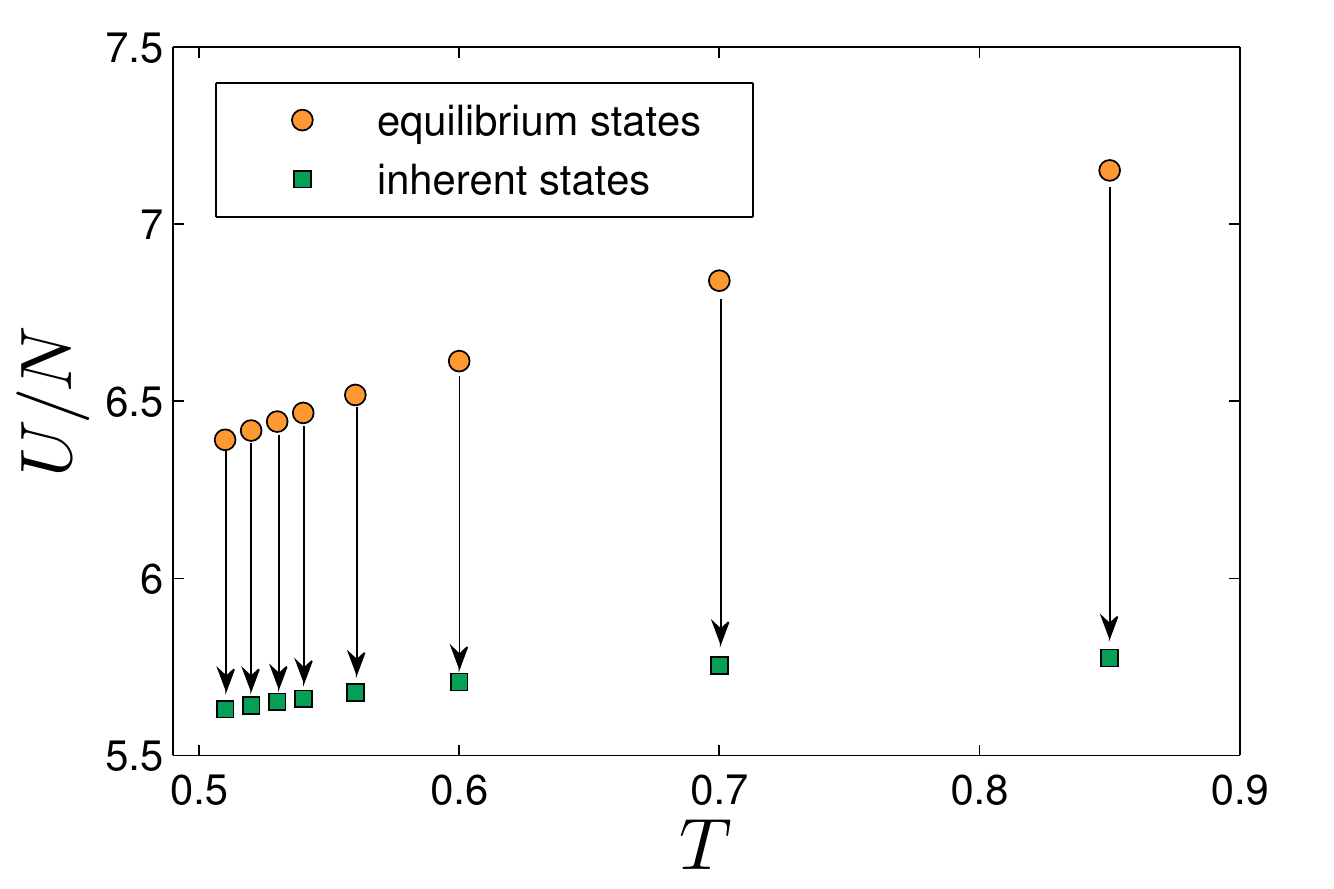}
\caption{\footnotesize Our ensembles of instantaneously quenched glassy samples were prepared by first generating equilibrium configurations at some temperature $T$. Each independent equilibrium configurations is quenched instantaneously to its underlying inherent state, as illustrated by the vertical arrows that connect between the pre- and post-quench sample-to-sample mean potential energy per particle.}
\label{parent_temp_ensemble_fig}
\end{figure}

We prepared two sets of ensembles of glassy samples. The first set in was created by performing an insantaneous quench, by means of a gradient descent minimization, of independent configurations of $N\!=\!2000$ particles in 3D, equilibrated at several different parent temperatures $T_0$, as illustrated in Fig.~\ref{parent_temp_ensemble_fig}. While these minimizations are computationally inefficient, they follow a physical dynamics which strongly suppresses inertial effects. Other minimization methods, e.g.~\cite{fire} could induce an uncontrolled degree of inertial effects, which we aimed at avoiding. Each ensemble corresponding to a particular parent temperature $T_0$ consists of 10000 independent glassy samples created as described above.

The second set of ensembles of glassy samples was created by performing a continuous quench at a finite quench rate $\dot{T}$, starting from equilibrium configurations at a temperature $T$. We created 3 such ensembles of systems of $N\!=\!2000$ particles, quenched at rates $\dot{T}\!=\!10^{-3}$, $\dot{T}\!=\!10^{-4}$, and $\dot{T}\!=\!10^{-5}$, starting from equilibrium configurations at temperatures $T\!=\!1.00$, $T\!=\!0.70$, and $T\!=\!0.51$, respectively. Each ensemble consists of 10000 independent glassy samples. To study finite-size effects, we have also created ensembles of glassy samples quenched at $\dot{T}\!=\!10^{-3}$, starting from equilibrium configurations at $T\!=\!1.00$ for system sizes ranging from $N\!=\!100$ to $N\!=\!1638400$ in 2D, and from $N\!=\!1000$ to $N\!=\!1000000$ in 3D. The number of independent glassy samples we created ranges from a few hundreds for our largest systems, to a few millions for our smallest systems. Temperature was controlled in our finite quench simulations by utilizing a Berendsen thermostat \cite{berendsen}, with a time parameter $\tau_{\mbox{\tiny Ber}}\!=\!4.0$ for $\dot{T}\!=\!10^{-3}$, and $\tau_{\mbox{\tiny Ber}}\!=\!10.0$ for $\dot{T}\!<\!10^{-3}$, see \cite{protocol_prerc} for a discussion about how this parameter is chosen.

\subsection{Supercooled liquid dynamics}

\begin{figure}[!ht]
\centering
\includegraphics[width = 0.50\textwidth]{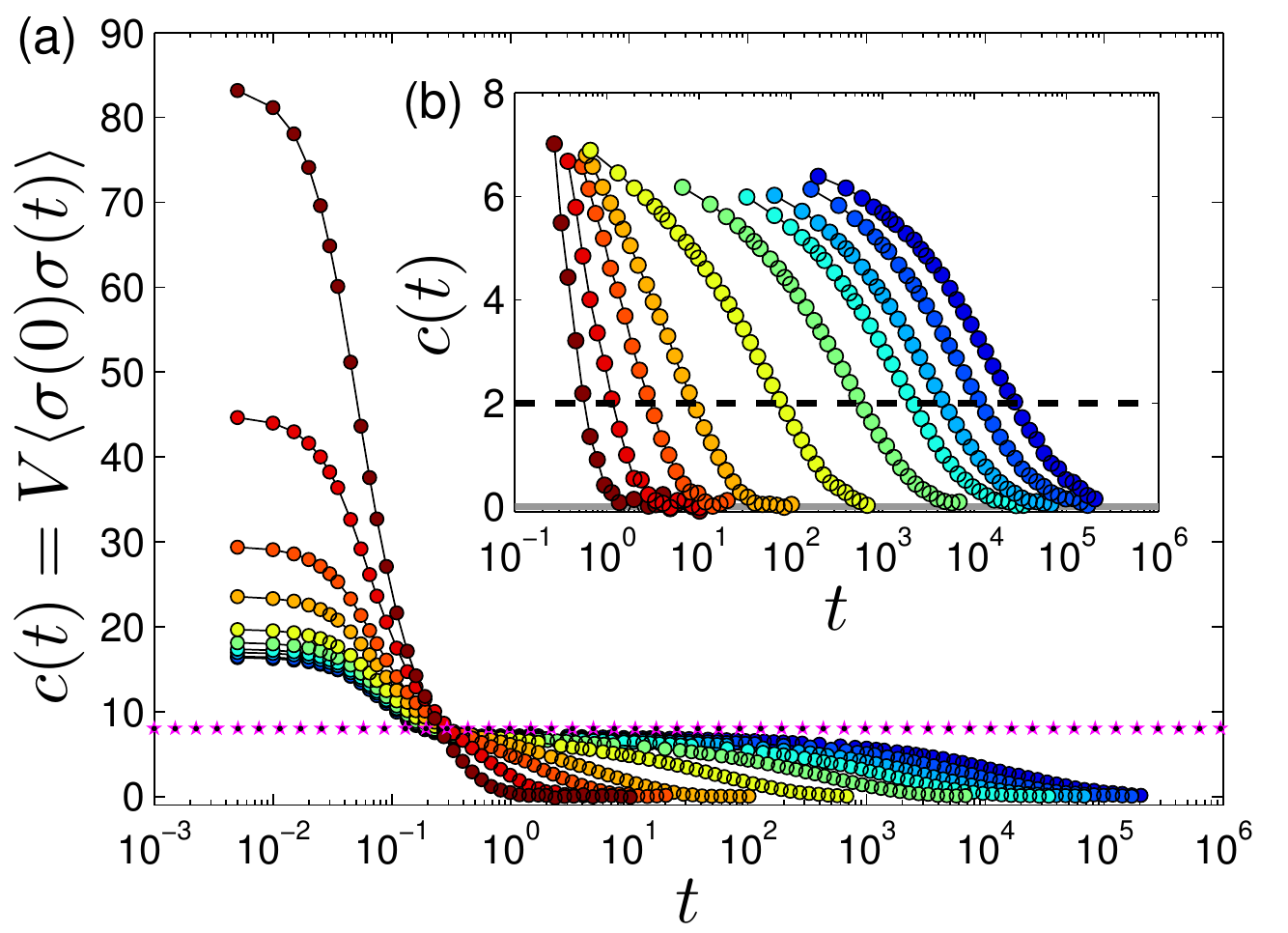}
\caption{\footnotesize Panels (a) and (b) show the full stress autocorrelation function $c(t)\!\equiv\!V\langle\sigma(0)\sigma(t)\rangle$ and the time-truncated stress correlation function, respectively, for temperatures $T\!=\!2.00,1.20, 0.85, 0.70, 0.60, 0.56, 0.54, 0.53, 0.52$ and $0.51$, decreasing from left to right in Panel~(b). The starred-line in Panel~(a) marks the onset of the two-step relaxation plateau, above which $c(t)$ is truncated in Panel (b). Relaxation times $\tau_\alpha$ (reported in Fig.~\ref{alpha_relaxation_fig}a) are calculated by estimating the time at which $c(\tau_\alpha)\!=\!2.0$, as indicated by the horizontal dashed line in Panel (b).}
\label{stress_correlation_fig}
\end{figure}

To quantify the relaxational dynamics of the supercooled liquid (in 3D), we monitored the stress autocorrelation function $c(t)\!\equiv\!V\langle\sigma(0)\sigma(t)\rangle$, displayed for various equilibrium temperatures in Fig.~\ref{stress_correlation_fig}.  Here $\sigma\!\equiv\!(\partial U/\partial\gamma)/V$, i.e.~we do not consider the kinetic contribution to the stress. The advantages of utilizing this correlation function are $(i)$ it requires no choice of parameters (compared e.g.~to the self-intermediate scattering function that requires chosing a wave vector, which could be temperature/protocol dependent), and (ii) the mean stress $\langle\sigma\rangle\!=\!0$ by symmetry. We estimate the primary structural relaxation time $\tau_\alpha$ by evaluating $c(\tau_\alpha)\!=\!2.0$, as shown in the figure. The lowest temperature that we are able to properly equilibrate our system is $T\!=\!0.51$. This leads us to the  operational determination of the computer glass transition temperature at $T_g\!=\!0.5$, which is used to rescale temperature axes in several figures throughout our work.

\subsection{Athermal quasistatic deformation}
\label{aqs}
In Fig.~\ref{shear_figure} we present data from athermal, quasistatic deformation of glassy sampes in 3D. These simulations are carried out by imposing small shear deformation increments, following each such increment with a standard nonlinear conjugate gradient minimization of the energy under the imposed shear. We employed Lees-Edwards periodic boundary conditions, and measured the stress-strain curves of glassy samples of $N\!=\!2000$ particles. These simulations were carried out on two sets of 2000 independent glassy samples, prepared using two protocols; the first protocol, referred to in the text and in the caption of Fig.~\ref{shear_figure} as the `fast quench', involves equilibrating liquid states at $T\!=\!2.00$, and instantaneously quenching each of those states to zero temperature using a gradient descent minimization. The second protocol, referred to in the text and in the caption of Fig.~\ref{shear_figure} as the `slow quench', involves equilibrating supercooled liquid states at $T\!=\!0.51$, and then carrying out a continuous quench to the glass of each independent equilibrium state at a rate $\dot{T}\!=\!10^{-5}$.

\subsection{Review of main observables}
\label{definitions}

We focus on several static and dynamic observables throughout this work. Vibrational modes were calculated by numerical diagonalization of the Hessian matrix ${\cal M}\!\equiv\!\frac{\partial^2U}{\partial\vec{x}\partial\vec{x}}$. Here, $\vec{x}$ denotes the vector of $N\!\times\!\dbar$ particles' coordinates in $\dbar$ dimensions, and $U\!=\!\sum_\alpha \varphi_\alpha(r_\alpha)$ is the potential energy, given in our model by a sum over pairwise contributions $\varphi_\alpha$, with $r_\alpha$ the pairwise distance, and $\alpha$ labeling pairs of interacting particles. All particles are assumed to share the same unit mass.

We also measured the athermal shear and bulk moduli of our glassy systems, given by \cite{lutsko}
\begin{equation}
G = \frac{\frac{\partial^2U}{\partial\gamma^2} - \frac{\partial^2U}{\partial\gamma\partial\vec{x}}\cdot{\cal M}^{-1}\cdot\frac{\partial^2U}{\partial\vec{x}\partial\gamma}}{V}\,,
\end{equation}
and
\begin{equation}
K = \frac{\frac{\partial^2U}{\partial\eta^2} - \frac{\partial^2U}{\partial\eta\partial\vec{x}}\cdot{\cal M}^{-1}\cdot\frac{\partial^2U}{\partial\vec{x}\partial\eta}}{V\dbar^2} + \frac{\dbar-1}{\dbar}p \,,
\end{equation}
respectively. Here $p$ is the hydrostatic pressure, $V$ is the system's volume, and $\gamma,\eta$ are simple shear and expansive strains, respectively, that parametrize the 2D strain tensor
\begin{equation}
\epsilon = \frac{1}{2}\left( \begin{array}{cc}2\eta + \eta^2& \gamma  + \gamma\eta\\ \gamma  + \gamma\eta & 2\eta + \eta^2 + \gamma^2\end{array}\right)\,,
\end{equation}
with a trivial extension to 3D.



In Sect.~\ref{system_size_section} of the main text we report measurements of the energies ${\cal M}\!:\!\hat{\pi}_4\hat{\pi}_4$ of quasilocalized excitations $\hat{\pi}_4$, referred to as `quartic modes'. The latter are solutions to the equation
\begin{equation}
{\cal M}\!\cdot\!\hat{\pi}_4 = \frac{{\cal M}\!:\!\hat{\pi}_4\hat{\pi}_4}{\frac{\partial^4U}{\partial\vec{x}\partial\vec{x}\partial\vec{x}\partial\vec{x}}\!::\!\hat{\pi}_4\hat{\pi}_4\hat{\pi}_4\hat{\pi}_4}\frac{\partial^4U}{\partial\vec{x}\partial\vec{x}\partial\vec{x}\partial\vec{x}}\tripleCdot\hat{\pi}_4\hat{\pi}_4\hat{\pi}_4\,,
\end{equation}
where the symbols $:,\tripleCdot$ and $::$ denote double, triple and quadruple contrations, respectively. In \cite{SciPost2016} these objects are introduced, and their characteristics are discussed in length. We have calculated a single solution $\hat{\pi}_4$ in each of our 2D glassy samples, as follows. First, the vibrational mode $\hat{\Psi}_{\mbox{\tiny min}}$ associated to the lowest vibrational frequency $\omega_{\mbox{\tiny min}}$ in a given sample is found by a conventional partial diagonalization of the Hessian matrix ${\cal M}\!\equiv\!\frac{\partial^2U}{\partial\vec{x}\partial\vec{x}}$. Then, $\hat{\Psi}_{\mbox{\tiny min}}$ is used as the initial conditions for a nonlinear minimization over directions $\vec{z}$ of the cost function
\begin{equation}
{\cal G}(\vec{z}) \equiv \frac{ ({\cal M}\!:\!\vec{z}\vec{z})^2 }{\frac{\partial^4U}{\partial\vec{x}\partial\vec{x}\partial\vec{x}\partial\vec{x}}\!::\!\vec{z}\vec{z}\vec{z}\vec{z}}\,.
\end{equation}
The cost function ${\cal G}$ assumes local minima at $\vec{z}\!=\!\hat{\pi}_4$, which is understood by realizing that $\frac{\partial {\cal G}}{\partial\vec{z}}\big|_{\hat{\pi}_4}\!=\!0$.

\section{structure-dynamics relations in glassy materials}
\label{sec:structure_dynamics}

In this Appendix present data from our own computer experiments demonstrating two widespreadly-known scenarios in which glassy dynamics features huge variations that are only accompanied by minor changes in conventional micro- and macrostructural measures.
\begin{figure}[!ht]
\centering
\includegraphics[width = 0.50\textwidth]{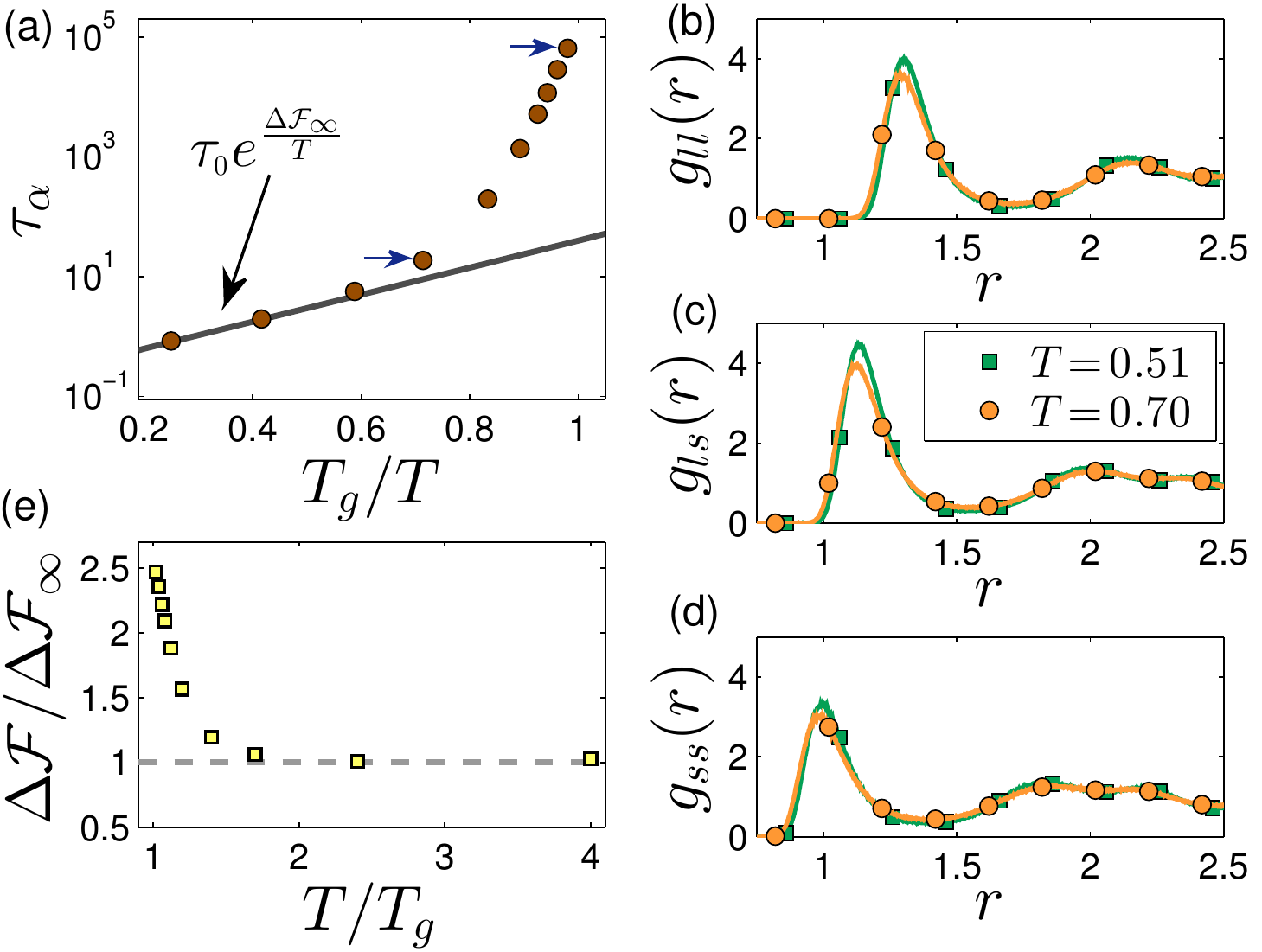}
\caption{\footnotesize (a) Equilibrium structural ($\alpha$) relaxation times for a model supercooled liquid (see Appendix~\ref{appendix} for details), vs.~$T_g/T$, where $T_g\!\approx\!0.5$ is the computer glass transition temperature of our model. The continuous line represents the high-temperature Arrhenius law $\tau_\alpha\!=\!\tau_0\exp(\Delta{\cal F}_\infty/T)$.  (e) Activation barriers $\Delta{\cal F}\!\equiv\!T\log(\tau_\alpha/\tau_0)$ normalized by $\Delta {\cal F}_\infty$ as extracted from the relaxation times of panel (a), vs.~temperature. (b)-(d) pair correlation functions $g(r)$ for large-large, large-small, and small-small pairs of particles, calculated in equilibrium configurations at $T\!=\!0.70$ and $T\!=\!0.51\!\approx\!T_g$, marked by the horizontal small arrows in panel (a).}
\label{alpha_relaxation_fig}
\end{figure}

In Fig.~\ref{alpha_relaxation_fig}a we plot the primary equilibrium structural relaxation time $\tau_\alpha$ of a generic supercooled glass forming model vs.~temperature. We observe the usual non-Arrhenius slowing down of dynamics until the computer glass transition temperature is reached. Fig.~\ref{alpha_relaxation_fig}b shows the relative increase in free-energy activation barriers, as deduced from the relaxation times (see figure caption for details). We observe a relative increase of 150\% in the free-energy activation barriers, while at the same time conventional structural variations, as seen in e.g.~pairwise spatial correlations displayed in panels (b)-(d), are very minor.

\begin{figure}[!ht]
\centering
\includegraphics[width = 0.50\textwidth]{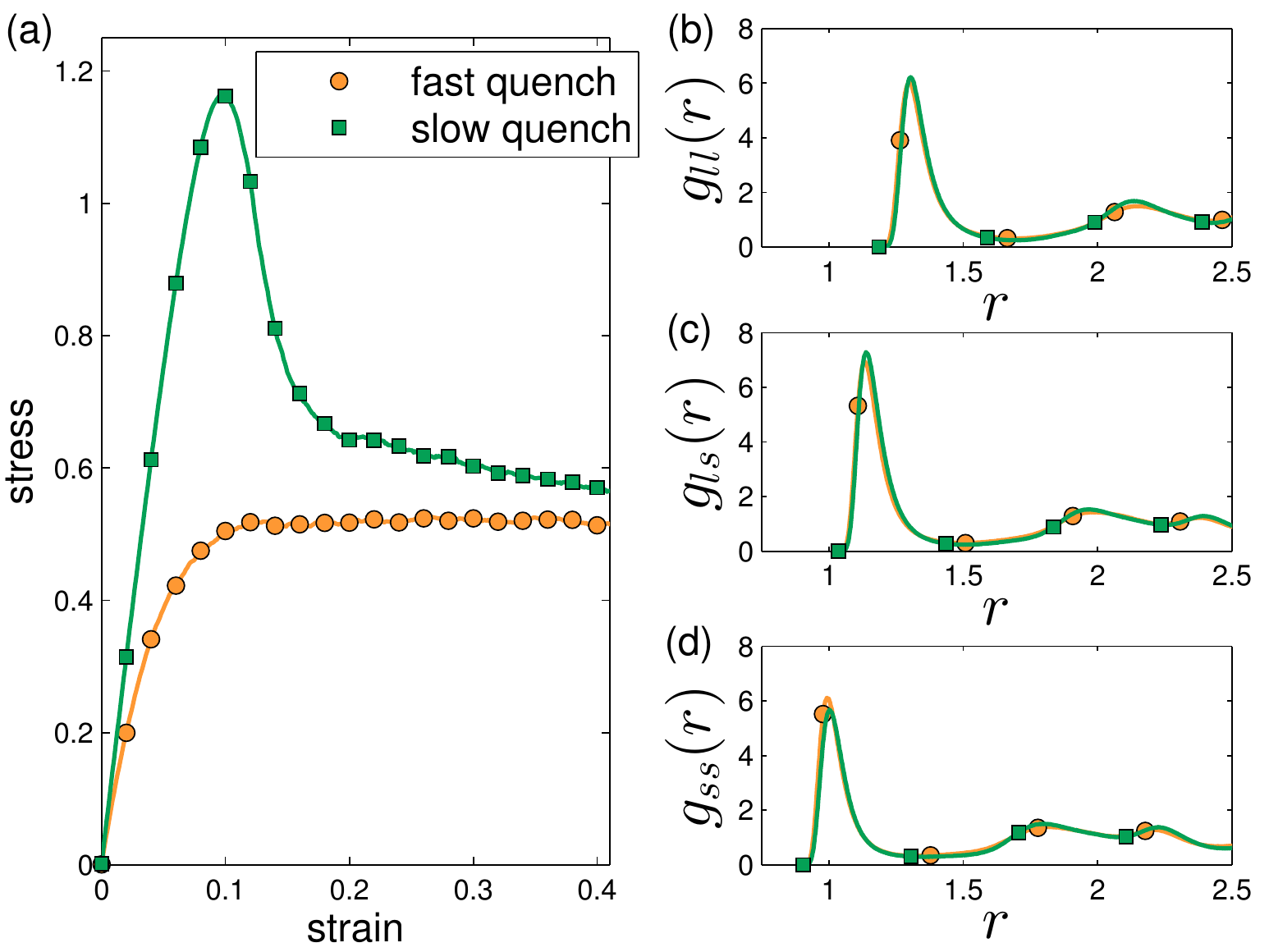}
\caption{\footnotesize (a) Stress-strain curves generated using an athermal, quasistatic protocol, averaged over for 2000 independent runs, for our ensembles of slowly-quenched (squares) and quickly-quenched (circles) glassy samples, see Appendix \ref{appendix} for details about the numerics. (b)-(d) pair correlation functions $g(r)$ for large-large, large-small, and small-small pairs of particles, calculated in both quickly quenched and slowly quenched ensembles.}
\label{shear_figure}
\end{figure}

Another example of enormous variations in dynamical responses accompanied by minor structural changes is illustrated in Fig.~\ref{shear_figure}a, where we plot stress-strain curves measured in computer experiments of athermal, quasistatic deformation of glassy samples, see Appendix~\ref{aqs} for details about the numerics. We show curves averaged over several independent realizations, starting from two ensembles of glassy samples that were quenched from the liquid phase at different rates (`fast quench' and `slow quench' in the legend, see Appendix~\ref{ensembles} for precise details about these preparation protocols). The difference in the mechanical response displayed by these two ensembles is impressive, given the insignificant variation in the their structure, as reflected by the pair correlation functions shown in panels (b)-(d) of Fig.~\ref{shear_figure}.

\begin{figure}[!ht]
\centering
\vspace{0.5cm}
\includegraphics[width = 0.50\textwidth]{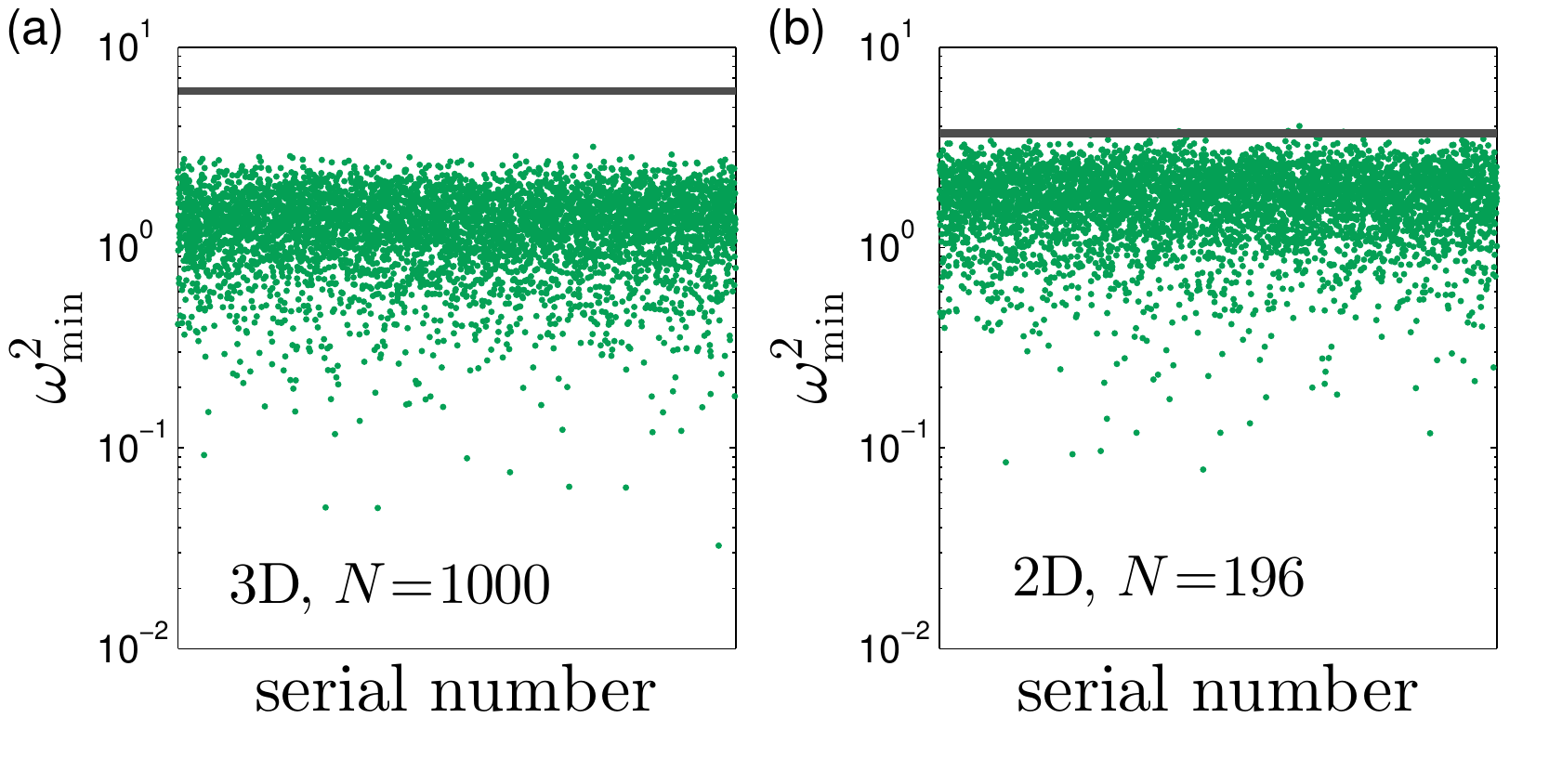}
\caption{\footnotesize Minimal vibrational frequencies squared measured in our glassy samples in (a) 3D and (b) 2D. The horizontal lines mark the energy of the lowest-frequency phonon, which notably bounds the lowest vibrational frequency's energy from above in 2D, even for very small systems as those employed for this test, but not in 3D.}
\label{bad_QLEs_fig}
\end{figure}

\section{Lowest-frequency vibrational modes\\ in 2D and 3D}
\label{sec:2D}

One of the main results established in this work is that the characteristic glassy energy scale as defined by the response to a local force dipole represents the energies of the softest quasilocalized excitations in the material. This relation is established by systematically changing some property of our glassy samples, observing the consequential induced relative variations in both the characteristic glassy energy scale and in the energy of the softest quasilocalized excitations, and establishing the degree of correlation between these induced relative variations. In Sect.~\ref{protocol_section} we followed this scheme to study the effect of varying the preparation protocol with which glassy samples were created, while in Sects.~\ref{sec:connection}-\ref{system_size_section} we induced variations in the aforementioned energy scales by systematically changing the size of the analyzed glassy samples, keeping the preparation protocol fixed.

Establishing a relation between the characteristic glassy energy and the energy of the softest quasilocalized excitations requires devising ways to robustly measure the latter. In 3D, as long as the system size is small enough (see elaborate discussion on this matter in \cite{modes_prl}), the softest quasilocalized excitations usually assume the form of harmomic vibrational modes. This was demonstrated in \cite{modes_prl}, and is further established in Fig.~\ref{bad_QLEs_fig}a, where we scatter-plotted the energies of the softest vibrational harmonic modes of glassy samples of size $N\!=\!1000$. The continuous horizontal line marks the energy of the lowest frequency phonons, as directly extracted from the system size and the measured shear moduli. It is clear that the energies of the lowest vibrational frequencies are well-separated from the energy of the lowest frequency phonon. This means that the lowest-frequency vibrational modes are not contaminated by the proximity of phonons with similar energies, and are therefore good representative of the softest quasilocalized excitations.

In 2D, the situation is dramatically different; we find that the energy separation between the lowest-frequency phonons and the lowest-frequency vibrational modes is very small, even for systems as small as $N\!=\!196$, as shown in Fig.~\ref{bad_QLEs_fig}b. In fact, the energy of the lowest-frequency phonon appears to \emph{bound} the energy of the lowest-frequency vibrational modes from above. Clearly, this bound will increasingly affect the energies of the lowest-frequency vibrational modes in larger systems.  We conclude that the lowest-frequency vibrational modes are not good representatives of soft quasilocalized excitations in 2D. It is therefore necessary to resort to an alternative approach to measure the energies of soft quasilocalized excitations; we opted for calculating the energies of quartic modes \cite{SciPost2016}, as described in Appendix~\ref{definitions}.

\end{document}